\documentstyle[12pt]{article}
\input psfig.sty
\def\nn{\noindent}
\def\Re{{\cal R \mskip-4mu \lower.1ex \hbox{\it e}\,}}
\def\Im{{\cal I \mskip-5mu \lower.1ex \hbox{\it m}\,}}
\def\ie{{\it i.e.}}
\def\eg{{\it e.g.}}

\def\etal{{\it et al.}}

\def\sub#1{_{\lower.25ex\hbox{$\scriptstyle#1$}}}
\def\tev{\,{\ifmmode\mathrm {TeV}\else TeV\fi}}
\def\gev{\,{\ifmmode\mathrm {GeV}\else GeV\fi}}
\def\mev{\,{\ifmmode\mathrm {MeV}\else MeV\fi}}
\def\to{\rightarrow}

\def\subw{_{\rm w}}
\def\mh{\ifmmode m\sbl H \else $m\sbl H$\fi}
\def\mch{\ifmmode m_{H^\pm} \else $m_{H^\pm}$\fi}
\def\mt{\ifmmode m_t\else $m_t$\fi}
\def\mc{\ifmmode m_c\else $m_c$\fi}
\def\mz{\ifmmode M_Z\else $M_Z$\fi}
\def\mw{\ifmmode M_W\else $M_W$\fi}
\def\mws{\ifmmode M_W^2 \else $M_W^2$\fi}
\def\mhs{\ifmmode m_H^2 \else $m_H^2$\fi}   
\def\mzs{\ifmmode M_Z^2 \else $M_Z^2$\fi}
\def\mts{\ifmmode m_t^2 \else $m_t^2$\fi}
\def\mcs{\ifmmode m_c^2 \else $m_c^2$\fi}
\def\mchs{\ifmmode m_{H^\pm}^2 \else $m_{H^\pm}^2$\fi}
\def\ztwo{\ifmmode Z_2\else $Z_2$\fi}
\def\zone{\ifmmode Z_1\else $Z_1$\fi}
\def\mtwo{\ifmmode M_2\else $M_2$\fi}
\def\mone{\ifmmode M_1\else $M_1$\fi}
\def\tb{\ifmmode \tan\beta \else $\tan\beta$\fi}
\def\xw{\ifmmode x\subw\else $x\subw$\fi}
\def\ch{\ifmmode H^\pm \else $H^\pm$\fi}
\def\lum{\ifmmode {\cal L}\else ${\cal L}$\fi}
\def\inpb{\,{\ifmmode {\mathrm {pb}}^{-1}\else ${\mathrm {pb}}^{-1}$\fi}}
\def\infb{\,{\ifmmode {\mathrm {fb}}^{-1}\else ${\mathrm {fb}}^{-1}$\fi}}
\def\epem{\ifmmode e^+e^-\else $e^+e^-$\fi}
\def\ppb{\ifmmode \bar pp\else $\bar pp$\fi}
\def\bsg{\ifmmode B\to X_s\gamma\else $B\to X_s\gamma$\fi}
\def\bsll{\ifmmode B\to X_s\ell^+\ell^-\else $B\to X_s\ell^+\ell^-$\fi}
\def\bstt{\ifmmode B\to X_s\tau^+\tau^-\else $B\to X_s\tau^+\tau^-$\fi}
\def\lamt{\ifmmode \tilde\lambda\else $\tilde\lambda$\fi}
\def\shat{\ifmmode \hat s\else $\hat s$\fi}
\def\that{\ifmmode \hat t\else $\hat t$\fi}
\def\uhat{\ifmmode \hat u\else $\hat u$\fi}

\newskip\zatskip \zatskip=0pt plus0pt minus0pt
\def\matth{\mathsurround=0pt}

\def\gsim{\mathrel{\mathpalette\atversim>}}
\def\atversim#1#2{\lower0.7ex\vbox{\baselineskip\zatskip\lineskip\zatskip
  \lineskiplimit 0pt\ialign{$\matth#1\hfil##\hfil$\crcr#2\crcr\sim\crcr}}}

\renewcommand{\thefootnote}{\fnsymbol{footnote}}

\begin{document} \begin{titlepage} 
\rightline{\vbox{\halign{&#\hfil\cr
&SLAC-PUB-7549\cr
&November 1997\cr}}}
\begin{center}

{\Large\bf Don't Stop Thinking About Leptoquarks: Constructing New Models}
\footnote{Work supported by the Department of 
Energy, Contract DE-AC03-76SF00515}
\medskip

\normalsize 
{\large JoAnne L. Hewett and Thomas G. Rizzo } \\
\vskip .3cm
Stanford Linear Accelerator Center \\
Stanford CA 94309, USA\\
\vskip .3cm

\end{center}

\begin{abstract} 

We discuss the general framework for the construction of new models containing 
a single, fermion number zero scalar leptoquark of mass $\simeq 200-220$ GeV 
which can both satisfy the D0/CDF search constraints as well as low energy 
data, and can lead to both neutral and charged current-like final states at 
HERA. The class of models of this kind necessarily contain new vector-like 
fermions with masses at the TeV scale which mix with those of the Standard 
Model after symmetry breaking. In this paper we classify all models of this 
type and examine their phenomenological implications as well as their 
potential embedding into SUSY and non-SUSY GUT scenarios.  The general 
coupling parameter space allowed by low energy as well as collider data for 
these models is described and requires no fine-tuning of the parameters.

\end{abstract} 

%\vskip0.45in
%\begin{center}

%Submitted to Physical Review {\bf D}.

%\end{center}

\renewcommand{\thefootnote}{\arabic{footnote}} \end{titlepage}

%%%%%%%%%%%%%%%%%%%%%%%%%%%%%%%---- Put text here
\section{Introduction and Overview}

\subsection{\it Current Status of the Leptoquark Scenario}

The observation of a possible excess of neutral current(NC) 
events in $e^+p$ collisions at high$-Q^2$ by both the H1{\cite {h1}} and 
ZEUS{\cite {zeus}} Collaborations have sparked much fervor in both the 
theoretical and experimental communities. This excitement has now been 
heightened by the recent announcement that both experiments may also 
be observing a corresponding excess in the charged-current(CC) 
channel{\cite {krak}}.  If these events are not merely a statistical 
fluctuation, it is clear that new physics must be invoked in order to provide a 
suitable explanation, \eg, compositeness appearing in the form of higher 
dimensional operators{\cite {vb}}, exotic modifications of the 
parton densities{\cite {parton}}, or the resonant production of a new 
particle{\cite {big,old}} such as a leptoquark (LQ) or squark in 
supersymmetric models with R-parity violation.  

If the excess is resonant in the $x$ distribution{\cite {web}},   
a popular interpretation{\cite {big,old}} 
invoked in the NC case is the $s-$channel production of a $\simeq 200-220$ GeV 
scalar (\ie, spin-0) leptoquark with fermion number ($F$) equal to zero. 
These quantum numbers arise from the requirements that ($i$) the observed 
excess appears in the $e^+p$ rather than the $e^-p$ channel, ($ii$) the 
Tevatron search constraints{\cite {tev}} exclude vector (spin-1) leptoquarks
with masses near 200 GeV, and ($iii$) any discussion of leptoquark models has 
been historically based on the classic work by Buchm\" uller, R\" uckl and 
Wyler (BRW){\cite {brw}}. In that paper the authors provide a set of 
assumptions under which consistent leptoquark models can be constructed; these 
we now state in a somewhat stronger form:

\begin{tabbing}
(a) \= LQ couplings must be invariant with respect to the Standard Model (SM)
gauge\\
\> interactions,\\
(b) \> LQ interactions must be renormalizable,\\
(c) \> LQs couple to only a single generation of SM fermions,\\
(d) \> LQ couplings to fermions are chiral,\\
(e) \> LQ couplings separately conserve Baryon and Lepton numbers,\\
(f) \> LQs only couple to the SM fermions and gauge bosons.
\end{tabbing}
Amongst these assumptions, both (a) and (b) are considered sacrosanct 
whereas (c)-(e) are data driven{\cite {rev}} by a host of low energy 
processes. Assumption (f) effectively requires that the leptoquark be the only 
new component added to the SM particle spectrum which seems quite unlikely in 
any realistic model. 
Based on these classical assumptions it is easy to show{\cite {brw}} that all 
$F=0$ scalar leptoquarks must have a unit branching fraction into a 
charged lepton plus jet (\ie, $B_\ell=1$). This lack of flexibility 
presents a new problem for the leptoquark interpretation of the HERA events 
for two reasons: ($i$) leptoquarks with $B_\ell=1$ clearly cannot 
accommodate any excess of events in the CC channel at HERA since these would 
require a sizeable leptoquark decay rate into neutrino plus jet,
($ii$) both CDF{\cite {cdf}} and D0{\cite {d0}} have recently presented 
new limits for the production of scalar leptoquarks at the Tevatron using the 
next-to-leading order cross 
section formulae of Kr\"amer \etal {\cite {kramer}}. In particular, in the
$eejj$ channel, D0 finds a $95\%$ CL lower limit on the mass of a $B_\ell=1$ 
first generation scalar leptoquark of 225 GeV. 
D0 has also performed a combined search for first generation leptoquarks by 
using the $eejj$, $e\nu jj$ and $\nu \nu jj$ channels. For fixed values of the 
leptoquark mass below 225 GeV, these search constraints can be used to place 
an upper limit on $B_\ell$. For $M_{LQ}$=200(210,220) GeV, D0 obtains the 
constraints $B_\ell \leq 0.40(0.62,0.84)$ at $95\%$ CL. Of course 
if CDF and D0 combine their searches in the future, then the 225 GeV bound
may rise to $\simeq 240$ GeV, in which case even 
stronger upper bounds on $B_\ell$ will be obtained.

Besides the obvious need to provide an potential explanation for the HERA data 
which satisfies all other experimental constraints, it is perhaps even more 
important to explore in a more general fashion how one can go beyond the rather 
restrictive BRW scenarios.  Even if the HERA events turn out to be 
statistical fluctuations, we will show that by the removal of the least 
tenable of the BRW assumptions, we can find important ways to extend the 
possible set of leptoquarks that may be realised in nature.
Since, as was mentioned above, it is difficult to believe that the addition of 
the leptoquark would be the only extension to the SM spectrum in any realistic 
model containing such a field, it is clear that assumption (f) should be
abandoned.  We now explore the consequences of this possibility.

\subsection{\it Enlarging the Framework of Leptoquark Models}

In order to satisfy all the experimental constraints it is clear that we 
need to have an $F=0$ scalar leptoquark as before, but now with a coupling to 
SM fermions given by
\begin{equation}
{\cal L}_{wanted} = [\lambda_u \nu u^c+\lambda_d ed^c]\cdot LQ +h.c.\,,
\end{equation}
with comparable values of the Yukawa couplings $\lambda_u$ and $\lambda_d$.  
This fixes the leptoquark's electric charge to be $Q_{LQ}=\pm 2/3$; no other 
charge assignment will allow the leptoquark to simultaneously couple 
to $ej$ and $\nu j$ as is required by the combination of 
HERA and Tevatron data. An alternative possibility, if neutrinos are Dirac
particles, or if
$\nu^c$ is light and appears as missing $p_T$ in a HERA or 
Tevatron detector, is the interaction 
\begin{equation}
{\cal L}_{wanted}' = [\lambda_u' \nu^c u+\lambda_d' e^cd]\cdot LQ' +h.c.\,
\end{equation}
It is important for later analysis to note that these two interactions 
cannot simultaneously exist as the BRW assumption (d) above would then be 
strongly violated.  Unfortunately, both of these 
Lagrangians as they stand violate assumption 
(a) above, in that they are not gauge invariant with respect to 
$SU(2)_L$.  We must then arrive at one of these effective interactions 
indirectly by some other means than by direct fundamental couplings.  In order 
to do so it is clear that we must be willing to abandon at least one of 
the BRW assumptions (a)-(f) and it is evident that (f) is the one most
easily dismissed.  Hence we will assume that the leptoquark has additional 
interactions besides those associated with SM gauge interactions and the 
Yukawa couplings to the SM fermions. We note, however, 
that fine-tuning solutions can be found which allow the assumption (c) 
to be dropped as a condition that applies in the mass eigenstate basis; 
these will not 
be discussed in detail here although it is important to understand how flavor 
mixing plays a role in leptoquark dynamics in realistic models.

In principle there are several alternatives as to what kinds of new additional 
interactions one can introduce, two of which we now briefly discuss. In a 
recent paper, Babu, Kolda and March-Russell{\cite {babu}} considered an 
interesting model with two different leptoquark doublets,
one coupling to $Ld^c$ and the other to $Lu^c$ (with 
$L$ being the SM lepton doublet). In this model the electric charge $Q=2/3$ 
members are mixed through a renormalizable coupling to the SM Higgs field
with the mixed leptoquarks forming mass 
eigenstates that can couple to both $ej$ and $\nu j$ as desired with the 
ratio of strengths controlled by the amount of mixing. The new interactions 
in this case are quite complex and a certain amount of fine 
tuning is necessary to get the spectrum and couplings to come out as desired. 
The rich phenomenology of this scenario, which now involves four leptoquark 
mass eigenstates of various charges, should be further studied in detail. 
A second scenario has only been briefly mentioned in the recent paper by 
Altarelli, Giudice and Mangano{\cite {altar}} who considered 
the possibility of at least temporarily violating both conditions 
(a) and (b) via non-renormalizable operators. These authors show, however, 
that both (a) and (b) can be restored by the introduction of new heavy 
fermions to which the leptoquarks couple in a gauge 
invariant fashion and which are then integrated out to obtain the desired 
effective low energy Lagrangian above. In this case there is only one
isosinglet $Q=2/3$ leptoquark present, which turns out to be quite 
advantageous. 

In this paper we will consider and classify all models wherein heavy fermions 
are used to generate the effective interactions ${\cal L}_{wanted}$ or 
${\cal L}_{wanted}'$ at low energies. 
As we will see, the emphasis of our approach is somewhat different than that
of Altarelli \etal, in that we will keep the new heavy fermions as active 
ingredients in our models and not treat them as an auxiliary device to produce 
the desired coupling structure.  In particular, we will assume that exotic, 
vector-like fermions exist and that the desired interactions are induced
through their couplings to the leptoquark and their mixing with the SM fermions.
The mixing between the new fermions and those of the SM will be generated
by conventional spontaneous symmetry breaking (SSB) via the usual 
Higgs doublet mechanism. It is 
only through SSB that the above effective Lagrangian can be obtained in 
the fermion mass eigenstate basis from an originally gauge invariant 
interaction. The small size of the effective Yukawa couplings in the 
above Lagrangians, ${\cal L}_{wanted}$ or ${\cal L}_{wanted}'$, 
are then directly explained by the same mechanism that produces the 
ordinary-exotic fermion mixing and automatically sets the scale of the 
vector-like fermion masses in the 
TeV region. We note that the use of vector-like fermions in 
this role is particularly suitable since in their unmixed state they make 
essentially no contribution to the oblique parameters{\cite {obl}}, they are 
automatically anomaly free, and they can have bare mass terms which are SM 
gauge invariant. (Alternatively, their masses can be generated by the vacuum 
expectation value of a SM singlet Higgs field.) 
Mixing with the SM fermions does not significantly detract from 
these advantages as we will see below. As is by now well-known{\cite {old}}, 
the leptoquark 
itself does not significantly contribute to the oblique parameters 
provided it is either an isosinglet, which will be the case realised in all 
of the models below, or in a degenerate multiplet.

\vspace*{-0.5cm}
\begin{figure}[htbp]
\centerline{
\psfig{figure=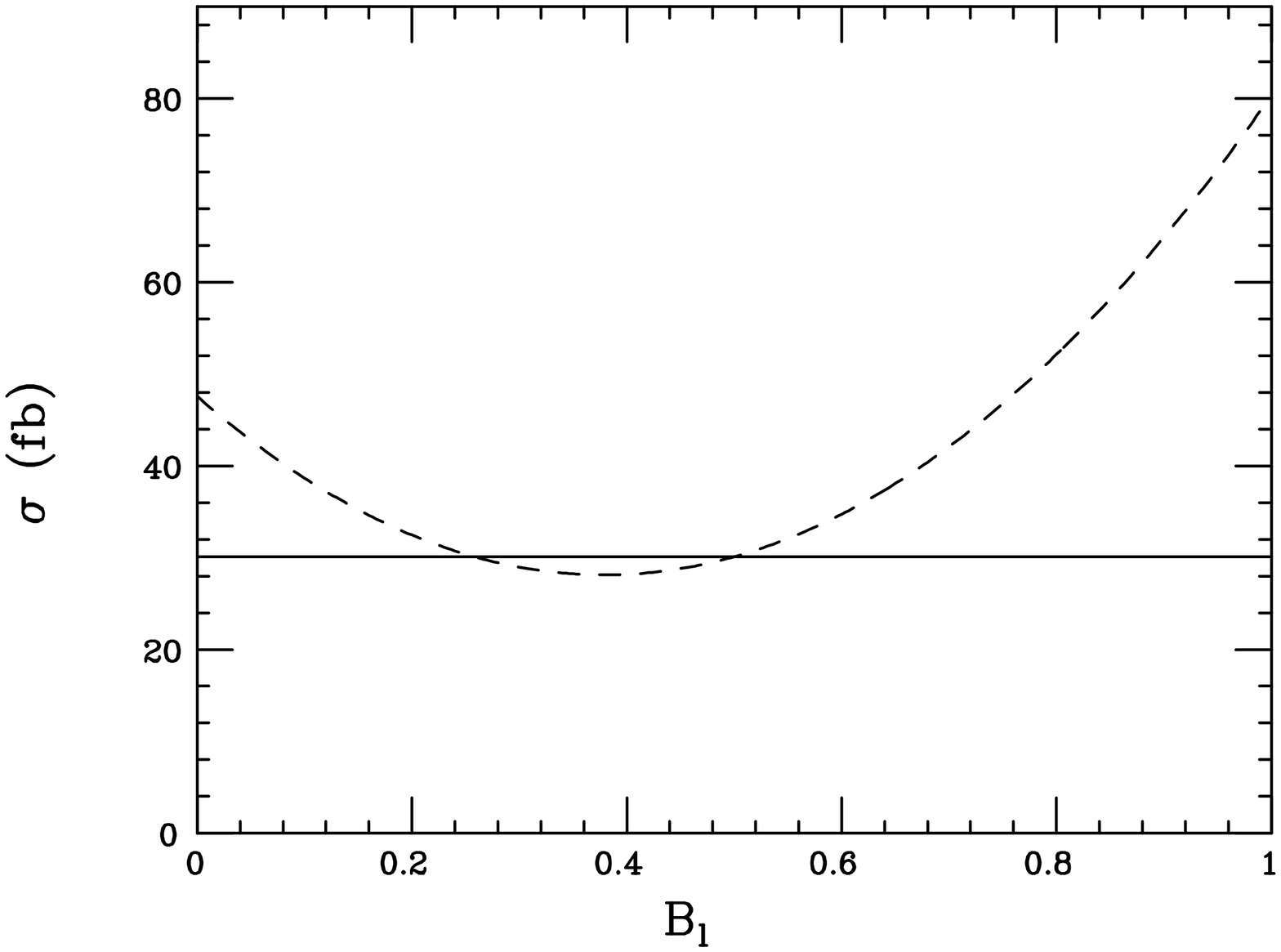,height=9.1cm,width=9.1cm,angle=-90}
\hspace*{-5mm}
\psfig{figure=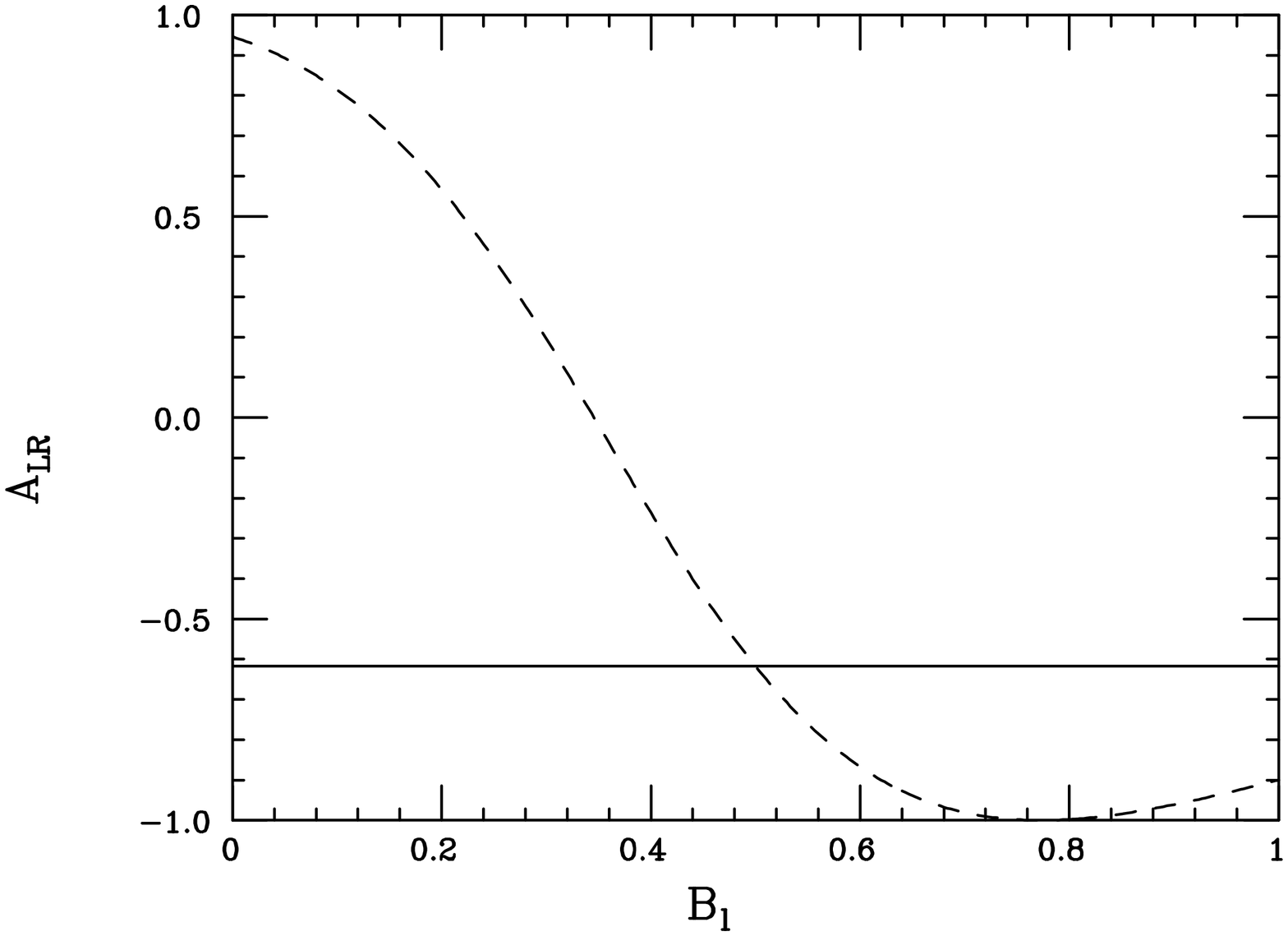,height=9.1cm,width=9.1cm,angle=-90}}
\vspace*{-1cm}
\caption{ Cross section(left) and associated polarization asymmetry(right) 
for the production of a pair of 200 GeV leptoquarks at a 500 GeV NLC. 
The dashed curve is the model of 
Babu, Kolda and March-Russell while the solid line is the prediction of the 
model with vector-like fermions.}
\label{sigmapol}
\end{figure}
\vspace*{0.4mm}

Before discussing the construction of new leptoquark models with vector-like 
fermions, it is interesting to note that HERA will not be 
able to distinguish between the two scenarios described above, 
even if the relative $ej$ and $\nu j$ branching fractions 
are precisely measured. The only means of differentiating the models is to 
either find the other new particles anticipated in each scheme, or to directly 
produce the $\simeq 200-220$ GeV leptoquarks at a high energy $e^+e^-$ 
collider such as the NLC{\cite {old}}. As we will see below, the 
charge and weak isospin of the leptoquark is fixed in the models with 
vector-like fermions and is independent of the value of $B_\ell$.  However, in 
the Babu \etal\ approach the leptoquark's effective weak isospin is highly 
correlated with the value of $B_\ell$. Fig.\ref{sigmapol} displays
a comparison of the leptoquark pair production cross section and 
polarization asymmetry for these two models at a 500 GeV NLC. It is clear
that unless $B_\ell$ is very close to $50\%$ the two scenarios will be 
easily separated at the NLC. These results also show that a leptoquark with the 
quantum numbers anticipated in vector-like fermion models 
is trivially distinguishable from the more conventional BRW leptoquarks by the 
same analysis{\cite {old}}.

\subsection{\it Constraints on Leptoquark Coupling Parameters}

\begin{figure}[htbp]
\centerline{
\psfig{figure=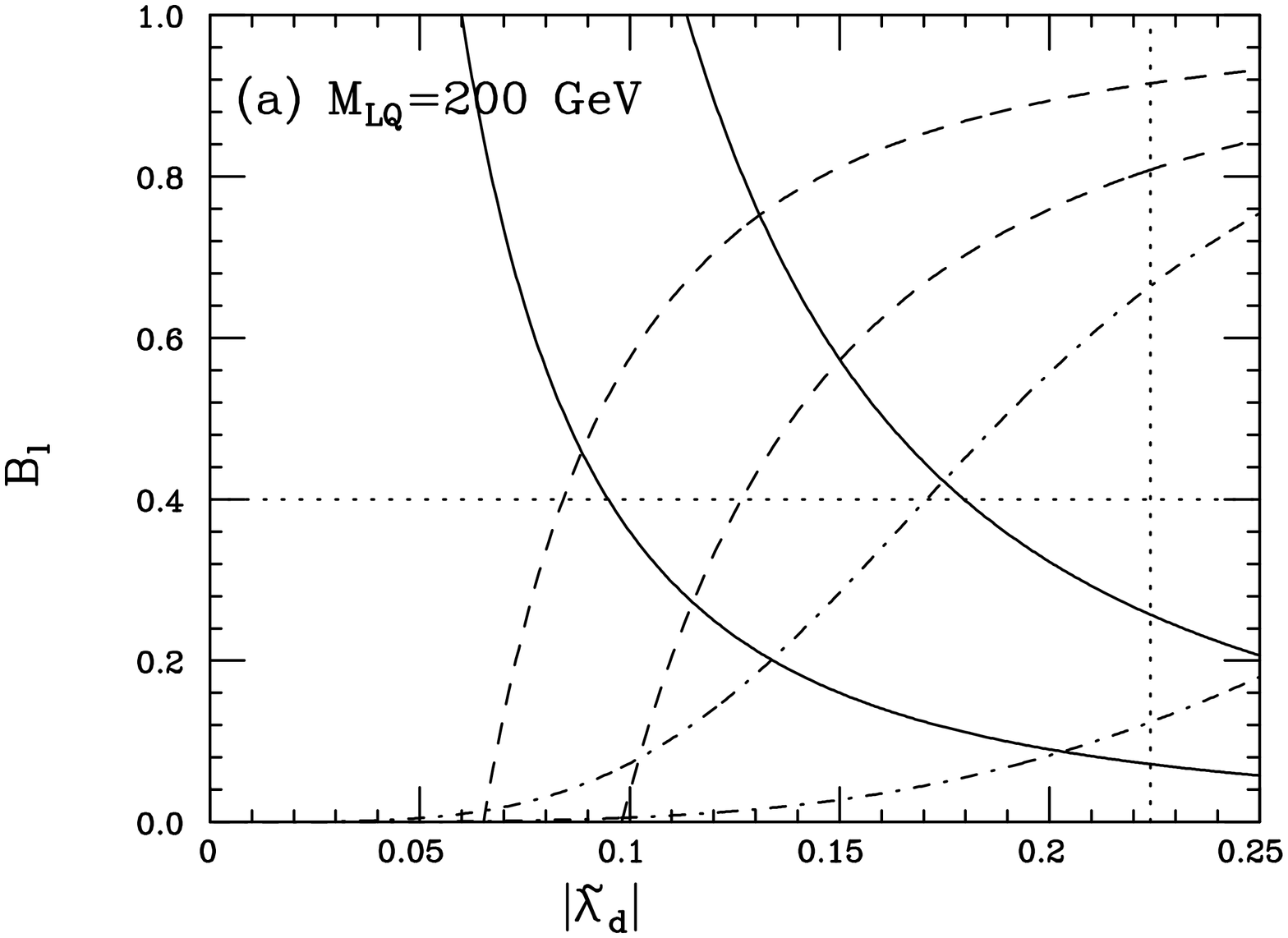,height=9.1cm,width=9.1cm,angle=-90}
\hspace*{-5mm}
\psfig{figure=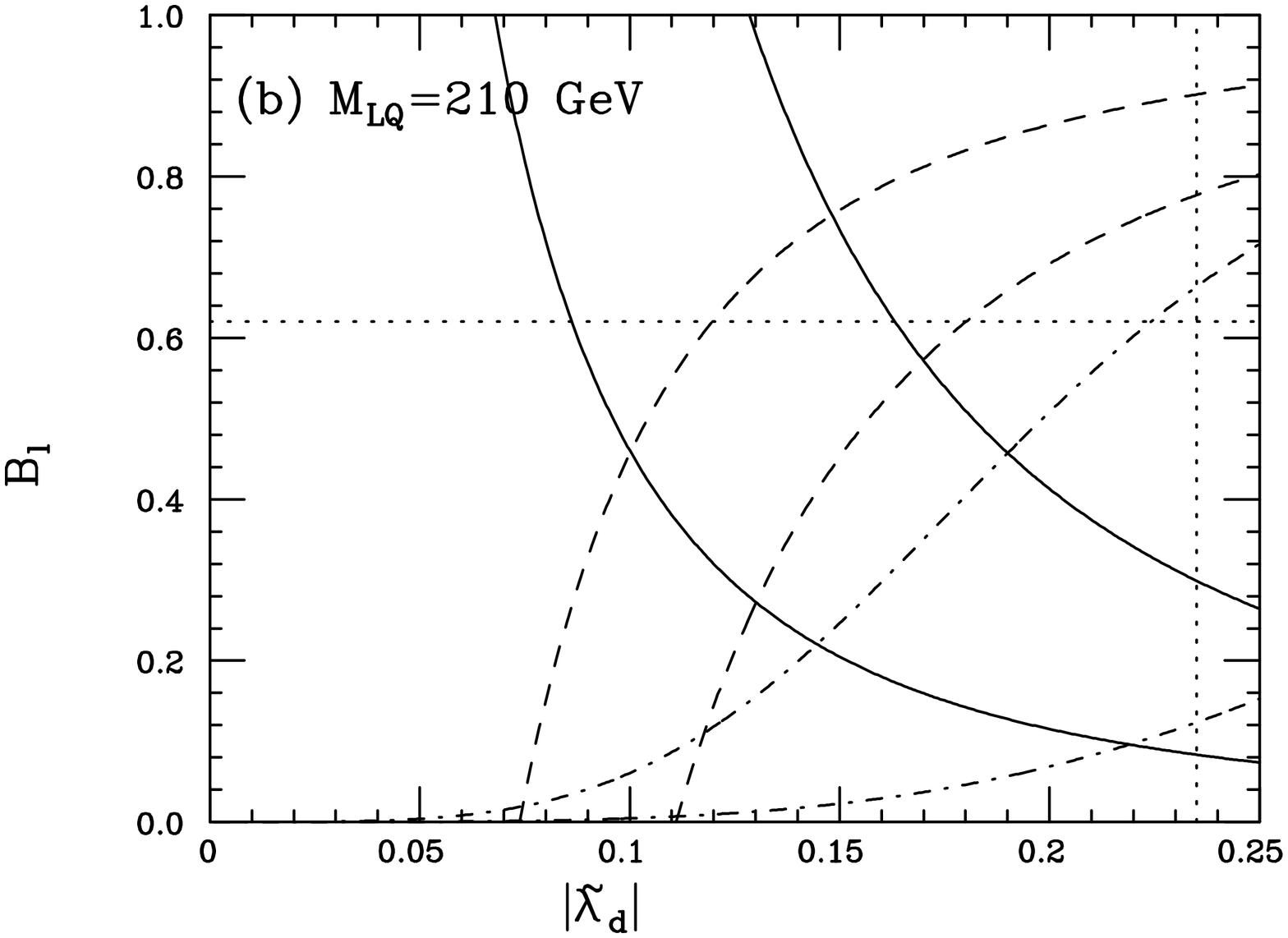,height=9.1cm,width=9.1cm,angle=-90}}
\vspace*{-0.75cm}
\centerline{
\psfig{figure=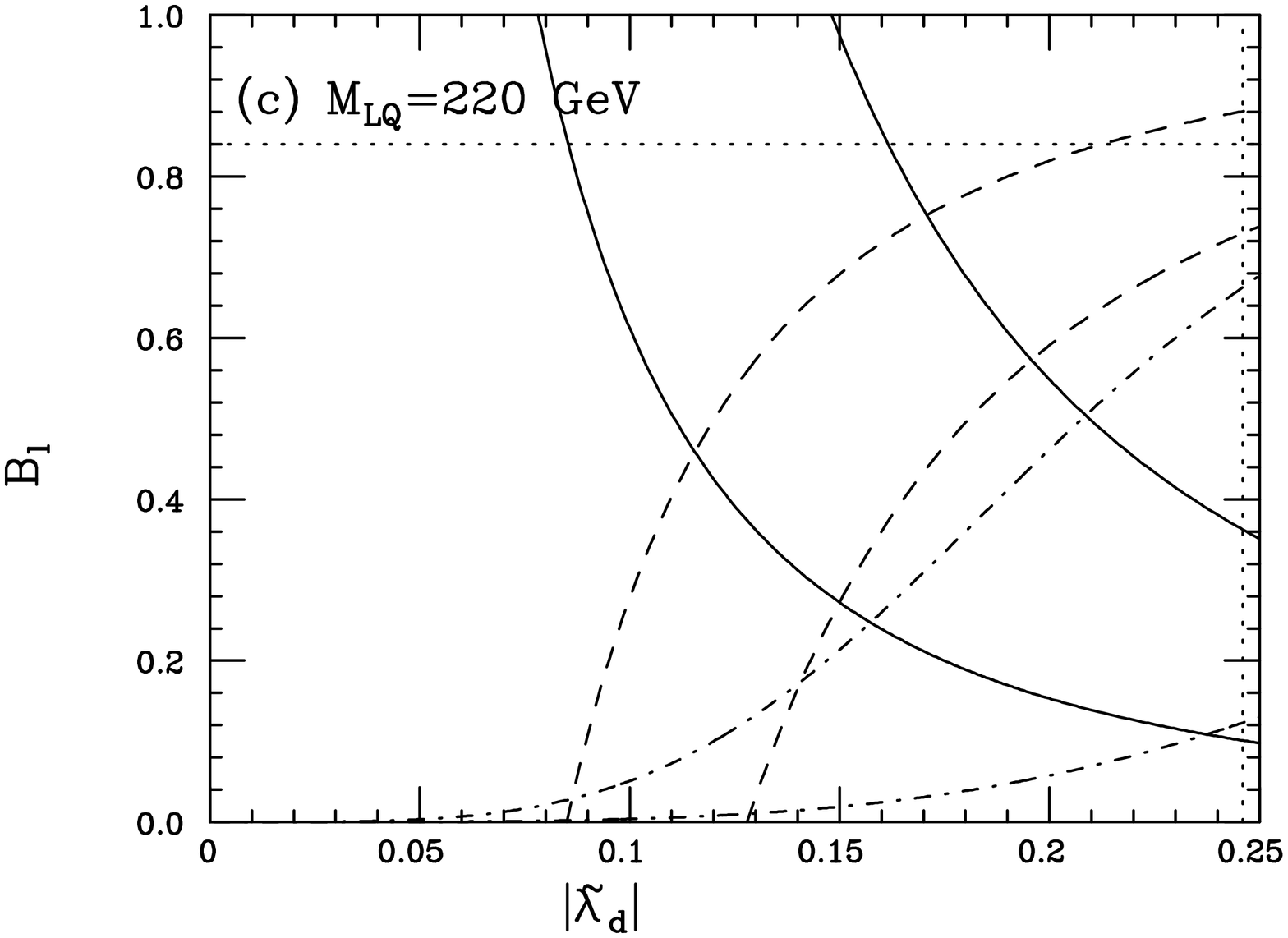,height=9.1cm,width=9.1cm,angle=-90}}
\vspace*{-1cm}
\caption{Allowed parameter space region in the $B_\ell-\tilde \lambda_d$ plane 
for a leptoquark with mass 200 GeV(top left), 210 GeV(top right) or 
220 GeV(bottom). 
The region allowed by Tevatron searches is below the horizontal dotted line 
while that allowed by APV data is to the left of the vertical dotted line. 
The region inside the solid band is required to explain the HERA excess in the 
NC channel at $1\sigma$.  The region between the dashed curves corresponds to
the $1\sigma$ range required to explain the apparent excess at HERA in the
charge current channel.  The region above the dash-dotted curve is allowed by 
$\pi$ decay universality: the lower(upper) curve corresponds to the case 
where $\tilde\lambda_u \tilde\lambda_d >(<)0$.}
\label{allow}
\end{figure}

As we will find below, in models with vector-like fermions, 
the only new physics at low energies introduced by the leptoquark itself 
can be parameterized in terms of the interactions in ${\cal L}_{wanted}$. 
It is then straightforward to use existing data to constrain the effective 
Yukawa couplings $\lambda_{u,d}$; here, we can express $\lambda_u$ in terms of
$B_\ell=\lambda_d^2/(\lambda_d^2+\lambda_u^2)$, since we assume that the 
leptoquark has no other decay modes. As discussed above, the 
Tevatron searches place a $\lambda_d$ independent constraint on $B_\ell$ 
for any fixed value of the leptoquark mass. Similarly, the recent measurements 
of Atomic Parity Violation (APV) in Cesium{\cite {wood}} place $B_\ell$ 
independent bounds on $\lambda_d${\cite {ros}} for fixed values of $M_{LQ}$. 
In addition, universality in $\pi$ decay 
constrains the product $\lambda_u \lambda_d${\cite {pdg}}, while the observed 
rate of NC events at HERA constrains instead 
the product $\lambda_d^2B_\ell$; in 
the later case QCD corrections are quite important{\cite {ks}}. The latest 
available results presented by both the ZEUS and H1 
Collaborations{\cite {straub}} in the neutral current as well as the charged 
current are included in our
estimate of the cross section constraints for both channels. (We note that
due to the relatively low statistics and other 
uncertainties the errors in this case are probably significantly 
underestimated so that this band is actually somewhat wider than what is 
shown below.) Combining these constraints defines an approximate allowed 
region in the $B_\ell-\tilde \lambda_d$ plane which is presented in 
Fig.\ref{allow} for $M_{LQ}=200, 210, 220$ GeV. Here, 
$\tilde \lambda=\lambda/e$ with $e$ being the conventional proton charge (this 
scaling of the coupling to $e$ follows earlier tradition{\cite {phyrep}}). We  
note that the size of the (apart from the HERA data) $95\%$ CL allowed region 
is sensitive to the two possible choices of the sign of the product of 
$\tilde \lambda_u \tilde \lambda_d$. As we will see below, the region 
corresponding to $\tilde \lambda_u \tilde \lambda_d>0$ is preferred so that 
the $\pi$ decay data has little impact in restricting the parameter 
space.  From Fig.\ref{allow} we see that the position of the allowed region 
moves up and to the right as the mass of the leptoquark increases from 200 
to 220 GeV. For the case $\tilde \lambda_u \tilde \lambda_d>0$, the size of 
the allowed region is not greatly affected as the leptoquark mass 
increases whereas, 
for $\tilde \lambda_u \tilde \lambda_d<0$ the region grows significantly in 
area with increasing mass. The size of the allowed region in each case would be 
substantially smaller if CDF and D0 could combine their results and further 
constrain the value of $B_\ell$.  

\vspace*{-0.5cm}
\nn

In addition to the constraints shown in Fig.\ref{allow}, further leptoquark 
coupling information can potentially be obtained\cite{altar} from examining 
the sum of the squares of the first row of the Cabbibo-Kobayashi-Maskawa (CKM)
weak mixing matrix, $\sum_i|V_{ui}|^2$.  
In the SM this sum is, of course, unity, but leptoquark exchange in $\beta$
decay can yield 
either an upward or downward shift in the extracted value of $|V_{ud}|$ of
\begin{equation}
|V_{ud}|^2_{eff}\simeq |V_{ud}|^2_{true}-1.52\times 10^{-3}\left( {200~GeV
\over M_{LQ} }\right)^2\left( {\tilde\lambda_u\over 0.15}\right)
\left( {\tilde\lambda_d\over 0.15}\right) \,,
\end{equation}
so that it would appear experimentally as if a unitarity violation 
were occurring. Interestingly, 
the value of the above sum has recently been discussed by Buras\cite{ajb},
who reports $\sum_i|V_{ui}|^2=0.9972\pm 0.0013$, which is more than $2\sigma$
below the SM expectation.  Clearly, if $\tilde\lambda_u\tilde\lambda_d>0$,
leptoquark exchange provides one possible additional contribution which, for
$\tilde\lambda_u=\tilde\lambda_d=0.15$ (implying $B_\ell=1/2$) 
and $M_{LQ}=200$ GeV,
would increase the sum to the value $0.9987$.  If we take 
this new determination of $V_{ud}$ seriously, then the constraint on leptoquark 
parameters from CKM unitarity can
be written in terms of $\tilde \lambda_d$ and $B_\ell$ at the $1\sigma$ level 
as (for the case of same sign leptoquark couplings) 
\begin{equation}
2.8\pm 1.3=1.52\left( {200~GeV\over M_{LQ}}\right)^2\left( {\tilde\lambda_d
\over 0.15}\right)^2\sqrt{ {1-B_\ell\over B_\ell}} \,,
\end{equation}
which is easily satisfied over most of the allowed parameter space in 
Fig.\ref{allow}.  As we will
see below, the mixing between the SM and vector-like fermions 
can also yield an additional small positive or negative contribution to 
$|V_{ud}|^2_{eff}$ which can have an effect on the CKM unitarity condition 
in some models.

\section{Analysis and Construction of New Leptoquark Models}

Employing the BRW assumptions (a)-(e) listed above we can construct our new 
extended set of leptoquark models using the following prescription:

($i$) The leptoquark couples a SM fermion multiplet, one of 
$(L,Q,u^c,d^c,e^c,(\nu^c))$, where $Q$ is the usual left-handed quark doublet, 
to an exotic vector-like fermion 
$X_i$ (or $X_i^c$) in a gauge invariant manner. For simplicity, the 
vector-like fermion is assumed to be an isosinglet or 
isodoublet under $SU(2)_L$ and either a singlet, triplet, or anti-triplet with 
respect to $SU(3)_C$. $X_i(X_i^c)$ will denote the new fields with fermion
number $F>(<)0$. 

($ii$) If $X_i$($X_i^c$) couples a SM fermion with a given helicity 
to the leptoquark, 
then $X_i^c$($X_i$) couples via $H$ or $H^c$ to the SM fermion of 
the opposite helicity, where $H/H^c$ are conventional doublet Higgs fields. We 
introduce both $H/H^c$ fields as independent degrees of freedom to allow for 
supersymmetrization of the models we construct. 

($iii$) To obtain the effective Lagrangian in Eqn. (1) we require that
terms of the form ${\cal N}{\cal U}^c+h.c.$ 
and ${\cal E}{\cal D}^c+h.c.$ must {\it both} appear in the original Lagrangian 
before spontaneous symmetry breaking by the 
$H/H^c$ vacuum expectation values (vevs),
where one of ${\cal N}({\cal U}^c)$ and ${\cal E}({\cal D}^c)$ must be a SM 
fermion field. This insures that a $Q=-1(0)$ lepton will couple to a 
$Q=-1/3(+2/3)$ quark to produce an $F=0$ leptoquark and 
that the type of structure in ${\cal L}_{wanted}$ 
can be obtained after mixing.  

($iv$) Bare mass terms for the fields $X_i$ of the form $M_iX_iX_i^c$ must be 
added to the original Lagrangian. 

($v$)  We follow the BRW assumptions (a)-(e) catalogued in the introduction.

We note that in the supersymmetric version of these models, the conjugate 
leptoquark field $LQ^c$ must also be present and that 
it cannot couple directly to any of the SM fermion fields, due to gauge 
invariance, unless it mixes with the leptoquark. This implies, in the zero 
$LQ-LQ^c$ mixing limit, that the conjugate leptoquark field cannot be 
produced at HERA, and that its production signature at the Tevatron will 
necessarily be quite different than that of the leptoquark and will have thus
escaped detection, even though the $LQ^c$ pair production cross section 
is the same as that for leptoquark pairs of the same mass.  We will briefly
discuss the more complex situation which includes this type of mixing below. 

We now begin to classify all possible models which employ SM and vector-like 
fermion mixing to obtain the desired leptoquark couplings.  We will take
one SM fermion multiplet at a time and pair it with a vector-like fermion and a 
leptoquark. Since there are six SM fermion multiplets (allowing for the 
possibility of $\nu^c$) there are naively at most six possible models that
can be constructed. (As we will see the actual number is somewhat more than 
this since various combinations of these models are feasible.)  To demonstrate 
how these construction rules work in practise, we begin by considering
the first case in detail.  Here, we couple an exotic fermion, denoted as $X_1$, 
to $L$ plus a leptoquark, \ie, $LX_1^c\cdot LQ$. In this case ($iii$) above
requires that $X_1$ be an isodoublet, with member charges of $2/3,-1/3$ since 
the leptoquark charge is fixed, as well as an $SU(3)_C$ triplet.  The BRW
assumption (a) then dictates that the leptoquark be an isosinglet. 
We can thus write $X_1^T=(U^0,D^0)$, where the superscript denotes
the  weak eigenstate fields. ($ii$) and ($iv$) above then instruct us to
add the SM gauge invariant terms $X_1u^cH+X_1d^cH^c$ and 
$M_1X_1X_1^c$ to the Lagrangian.  Including the Yukawa couplings (which we 
assume are of order unity) these terms, together with the gauge interactions 
of both the leptoquark and the fermion doublet $X_i$, form our new set of
interactions that are added to the SM.  Denoting this as
model A, we thus arrive at 
\begin{equation}
{\cal L}_A = \lambda_A LX_1^c\cdot LQ+a_uX_1u^cH+a_dX_1d^cH^c-M_1X_1X_1^c
+gauge+h.c.\,,
\end{equation}
where `$gauge$' represents the new gauge interactions of the leptoquark and 
$X_1$.  We emphasize that all of the above Yukawa couplings are assumed to be 
of order unity. 

When $H$ and $H^c$ receive vevs ($v$ and $v^c$), the $a_{u,d}$ terms in the
above Lagrangian induce off-diagonal couplings in both the $Q=-1/3$ and $Q=2/3$
quark mass matrices.  Neglecting the $u$- and $d$-quark masses, these are 
given in the $\bar \psi_L^0 M \psi_R^0$ weak eigenstate basis by
\begin{eqnarray}
\bar \psi_L^0 M_{u}\psi_R^0 & = & (\bar u^0,\bar U^0)_L \left( \begin{array}{cc}
0 & 0  \\
a_uv  & -M_1  
\end{array} \right)\left( \begin{array}{c}
u^0 \\
U^0
\end{array} \right)_R  \,, \\
\bar \psi_L^0 M_{d}\psi_R^0 & = &(\bar d^0,\bar D^0)_L  \left( \begin{array}{cc}
0 & 0  \\
a_dv^c  & -M_1  
\end{array} \right) \left( \begin{array}{c}
d^0 \\
D^0
\end{array} \right)_R  \,.
\end{eqnarray}
Both $M_{u,d}$ can be diagonalized by a bi-unitary transformation which 
becomes bi-orthogonal under the assumption that 
the elements of $M_{u,d}$ are real, resulting in the diagonal mass matrices
$M_{u,d}^{diag}=U_L(u,d)M_{u,d}U_R(u,d)^\dagger$. Since $U_{L,R}(u,d)$ are 
simple $2\times 2$ rotations they can each be 
parameterized by a single angle $\theta_{L,R}^{u,d}$. For the case at hand 
$\theta_L^{u,d}=0$, whereas $\theta_R^{u,d}\simeq a_{u,d}v(v^c)/M_1$.
% assuming that $M_1 >>v,v^c$. 
Taking the Yukawa couplings to be of order unity and $v,v^c\sim 100-250$ GeV,
the size of the mixing is 
fixed by the scale of $M_1$. Writing $U^0\simeq U+\theta_R^u u$ in terms of the
mass eigenstate fields, and similarly for $D^0$, the interaction involving the 
SM fermions and the leptoquark thus becomes
\begin{equation}
{\cal L}_{light} = \left[\left({\lambda_A a_u v\over M_1}\right)~\nu u^c
+\left({\lambda_A a_dv^c\over M_1}\right)~ed^c\right]\cdot LQ +h.c.\,,
\end{equation}
which is the exact form we desired in Eqn. (1).  This naturally leads to a 
reasonable relative branching fraction for the $LQ\to \nu j$ decay mode, and 
gives acceptable values for $\lambda_{u,d}$ in Eqn. (1) for $M_1$ in the 1-5 
TeV range. 

At this point one may note that we have omitted a term in ${\cal L}_A$ of the 
form $-M'QX_1^c$, with $M'$ being a bare mass parameter. 
Such a term is, of course, gauge invariant and should be present in principle 
but has little influence on the scenario as far as the leptoquark interactions
are concerned. Of course one can always invent a symmetry 
to forbid this term if so desired as in practice such a term may produce an 
uncomfortably large mass for the SM fermions, induced by mixing, and so 
additional care is required. However, to keep the following discussion as 
general as possible, such terms will be included in our discussion.
With $M'$ being the same order as $M_1$ there is essentially no change in 
our result for the right-handed mixing above; we now obtain 
$\theta_R^{u,d}\simeq a_{u,d}v(v^c)M_1/(M_1^2+M'^2)\simeq a_{u,d}v(v^c)/M_1$.
However, $M'$ induces a non-zero mixing for the left-handed fields, but this
does not influence either the 
leptoquark or $Z$ boson couplings to the light fermions.  There is a new
contribution in the case of the light fermions' charged current couplings to 
the $W$, but it is quite suppressed being proportional to 
$\Delta=1-\cos(\theta_L^u-\theta_L^d)$, with the difference 
$\theta_L^u-\theta_L^d\simeq (a_u^2v^2-a_d^2v^{c2})/M_1^2$ being small. 
Note that while both $\theta_L^{u,d}$ are large, neither is directly 
observable and it is the {\it difference} between the two, which is very 
small, that is observable. It is also important to remember that this mixing 
angle difference is also proportional 
to $B_\ell^2-(1-B_\ell)^2$, so that it is further suppressed for values of 
$B_\ell$ approaching 0.5.  Even without considering these cancellations, we
estimate the effect to be very small since the difference in the left-handed
mixing angles is roughly given by
$\theta_L^u-\theta_L^d \sim \theta_R^2 \simeq 
(0.05)^2$, implying $\Delta < 10^{-5}$. In the next section
we will return to the general question of whether the 
effects associated with the finite size of these mixing angles can lead to 
observable shifts in SM expectations.

To proceed with our systematic analysis, we first list the remaining five  
skeleton models that are obtained by simply combining the other SM 
representations with an appropriate vector-like fermion 
and leptoquark field (note that both models B 
and F involve the field $\nu^c$): 
\begin{eqnarray}
{\cal L}_B & = & \lambda_B QX_2^c\cdot LQ+a_eX_2e^cH^c+a_{\nu} X_2 \nu^cH
-M_2X_2X_2^c\,, \nonumber\\
{\cal L}_C & = & \lambda_C X_3u^c\cdot LQ+a_1LX_3^cH 
-M_3X_3X_3^c\,, \nonumber\\
{\cal L}_D & = & \lambda_D X_4d^c\cdot LQ+a_2LX_4^cH^c 
-M_4X_4X_4^c\,, \\
{\cal L}_E & = & \lambda_E X_5e^c\cdot LQ+a_3QX_5^cH^c
-M_5X_5X_5^c\,, \nonumber\\
{\cal L}_F & = & \lambda_F X_6\nu^c\cdot LQ+a_4QX_6^cH
-M_6X_6X_6^c\,, \nonumber
\end{eqnarray}
where the usual `{\it gauge + h.c.}' terms have been dropped for simplicity. 
Note that model B is essentially the leptonic equivalent of model 
A; here, the vector-like fermion 
field $X_2$ is a color singlet, weak isodoublet, \ie , $X_2^T=(N^0,E^0)$,
and the leptoquark remains an isosinglet. This model requires the 
neutrino to be a light Dirac field or, at the very least, $\nu^c$ to appear as 
missing $p_T$ in the leptoquark decay process. 

It is important to notice that some of 
these individual skeleton models {\it do not} 
satisfy all of the model building constraints listed above, in 
particular ($iii$). However, this requirement can be satisfied by taking 
combinations of the various skeleton ${\cal L}_i$ above, taking 
care not to violate the BRW condition (d) that the leptoquark 
couplings remain chiral.  The weaknesses in models 
C and D as well as E and F can be overcome by simply pairing them: 
\begin{eqnarray}
{\cal L}_{CD} & = & [\lambda_C Nu^c+\lambda_D Ed^c]\cdot LQ+a_1LN^cH 
+a_2LE^cH^c -M_NNN^c-M_EEE^c\,, \nonumber\\
{\cal L}_{EF} & = & [\lambda_E De^c+\lambda_F U\nu^c]\cdot LQ+a_3QD^cH^c
+a_4QU^cH-M_DDD^c-M_UUU^c\,, 
\end{eqnarray}
where the superscript `0' denoting the weak eigenstate has been dropped for 
simplicity. Both models CD and EF now satisfy all of our model building 
requirements, however, it is important to realize that these two combinations
are {\it not} the only set of alternatives.  In this case,
the fields $U,D,N$ and $E$ are identified with 
$X_{6,5,3,4}$, respectively, and are all weak isosinglets with the field 
notation designating the color and charge information.
Note that the individual bare mass terms are present for 
these fields, and that in both models the leptoquark is again an 
isosinglet with charge $2/3$. As in the case of model B, model EF 
requires $\nu^c$ to appear as missing $p_T$ in the leptoquark 
decay, otherwise these models are excluded by the apparent HERA CC excess and, 
possibly, by the Tevatron constraints. We note that, as in the case 
of model A, additional gauge invariant mass/mixing terms can be added to the 
Lagrangians of models B, CD and EF.  These take the form of $-M_L'LX_2^c$ 
for model B, $-M_N'N\nu^c-M_E'Ee^c$ for model CD and $-M_D'Dd^c-M_U'Uu^c$ for 
model EF.  They produce essentially no additional new physics 
effects at a visible level in the latter two cases since none of the SM fermion 
couplings to the gauge bosons are further altered.  However, in model B, as 
was seen for model A, a modification of the SM leptonic CC couplings to the 
$W$ boson will occur and is proportional to the 
square of the difference in the left-handed mixing angles needed to 
diagonalize the 
neutral and charged lepton mass matrices. For completeness, the mass matrices 
for the vector-like and SM fermion sector for each of these models are as 
follows (using the same weak eigenstate basis as above): for model B, 
\begin{eqnarray}
M_{\nu} & = & \left( \begin{array}{cc}
0 & -M_L'  \\
a_\nu v  & -M_2  
\end{array} \right) \,, \\
M_{e} & = & \left( \begin{array}{cc}
0 & -M_L'  \\
a_ev^c  & -M_2  
\end{array} \right) \,,
\end{eqnarray}
whereas for model CD we find 
\begin{eqnarray}
M_{\nu} & = & \left( \begin{array}{cc}
0 & a_1 v  \\
-M_N'  & -M_N  
\end{array} \right) \,, \\
M_{e} & = & \left( \begin{array}{cc}
0 & a_2v^c  \\
-M_E'  & -M_E 
\end{array} \right) \,,
\end{eqnarray}
and for model EF we correspondingly obtain 
\begin{eqnarray}
M_{u} & = & \left( \begin{array}{cc}
0 & a_4v  \\
-M_U'  & -M_U  
\end{array} \right) \,, \\
M_{d} & = & \left( \begin{array}{cc}
0 & a_3v^c   \\
-M_D'  & -M_D  
\end{array} \right) \,,
\end{eqnarray}
where in all cases we have allowed for the additional gauge invariant terms 
discussed above. With the primed and unprimed bare mass terms of roughly the 
same magnitude, the effective $\lambda_{u,d}$ (or $\lambda_{u,d}'$) couplings 
can be read off directly from these matrices and the above Lagrangians are 
expressible in the same form as in Eq. 8 with the appropriate substitutions of 
masses and Yukawa couplings. It is important to remember that in 
models CD and EF, which involve mixings with isosinglet fermions, the 
roles of the left- and right-handed mixing angles, $\theta_L$ and $\theta_R$, 
are essentially interchanged with respect to those in models A and B 
where the vector-like fermions are 
in isodoublets. In all cases the 
relevant mixing angles are of order 0.05 as obtained in the case of model A,
due to the phenomenological constraints imposed on the effective Yukawa 
couplings by the HERA data.

Lastly, we note that models A, B, CD and EF are not the only successful 
ones that can be constructed.  We can, \eg, take either model 
A or B and combine it with 
one of the skeleton models C-F; for example, model B could be coalesced
with F. In principle, many potential hybrid models of 
this type can be constructed. 
This observation will be important below when we discuss the 
unification of these models within a GUT framework, as well as the 
phenomenological implications 
of the SM and vector-like fermion mixing.
We note that in these more 
complex models the fermion mixing(s) that generate the SM fermion 
couplings to the leptoquark can arise from multiple sources. Of course,
when we attempt to construct further hybrid models, we must take care not to 
violate the assumption that the leptoquark 
couplings are chiral. Given this very strong 
constraint, the entire list of models that can be constructed in this fashion 
are only ten in number: A, B, CD, EF; AC, AD, ACD, BE, BF and BEF. We note 
that models A, CD, AC, AD and ACD produce the effective interaction 
${\cal L}_{wanted}$, 
while models B, EF, BE, BF and BEF produce instead ${\cal L}_{wanted}'$. 
The models and the exotic fermions associated with each of 
them are catalogued in Table 3.
\begin{table}
\centering
\begin{tabular}{|c|c|} \hline\hline
Model & Vector-like Fermions \\ \hline
 & \\
A & $\left( \begin{array}{c}
            U \\
            D
\end{array} \right)_{L,R}$ \\[4ex]
CD & $N_{L,R}; E_{L,R}$ \\[2ex]
AC & $\left( \begin{array}{c}
            U \\
            D
\end{array} \right)_{L,R}; N_{L,R}$ \\[4ex]
AD & $\left( \begin{array}{c}
            U \\
            D
\end{array} \right)_{L,R}; E_{L,R}$ \\[4ex]
ACD & $\left( \begin{array}{c}
            U \\
            D
\end{array} \right)_{L,R}; N_{L,R}; E_{L,R}$ \\[4ex]
\hline
 & \\
B & $\left( \begin{array}{c}
            N \\
            E
\end{array} \right)_{L,R}$ \\[4ex]
EF & $U_{L,R}; D_{L,R}$ \\[2ex]
BE & $\left( \begin{array}{c}
            N \\
            E
\end{array} \right)_{L,R}; D_{L,R}$ \\[4ex]
BF & $\left( \begin{array}{c}
            N \\
            E
\end{array} \right)_{L,R}; U_{L,R}$ \\[4ex]
BEF & $\left( \begin{array}{c}
            N \\
            E
\end{array} \right)_{L,R}; U_{L,R}; D_{L,R}$ \\[4ex]
\hline\hline
\end{tabular}
\caption{Listing of models and the vector-like fermions which
are contained in them.}
\end{table}

\section{Implications and Tests}

Some phenomenological implications of these models are examined in
this section.  The detailed phenomenology depends on whether or not 
supersymmetry (SUSY) is also introduced. Clearly, the non-SUSY 
versions are more easily analyzed but both classes of models share many common 
features which we will discuss here. These include new interactions due to
the mixing between the vector-like and SM fermions as well as from the 
existence of the leptoquark and the vector-like fermions themselves. 

\subsection{\it Direct Production of Vector-Like Fermions}

The production and decay of vector-like fermions has been extensively
discussed in the literature{\cite {phyrep,dpf}}, particularly in the context
of $E_6$ grand unified theories.  The mixing induced between these new fields
and the ordinary SM fermions not only modifies the SM fermion couplings to 
the $W$ and $Z$ but also leads to flavor-changing $Z$ interactions involving 
a single SM fermion and a vector-like fermion.  This implies that the 
vector-like fermions can be produced in pairs via the usual mechanisms, or
singly via mixing.  Once produced, they can decay through 
mixing into a SM fermion and a $Z$ or $W$ with comparable rates.
However, unlike most models 
containing vector-like fermions, it is more likely here that at least some of 
these states will dominantly decay to leptoquarks instead due to the large 
assumed size of the Yukawa couplings.  Given the expected large mass of the
new fermions in these models, they will only be accessible at the LHC
(until $\sqrt s$=2-10 TeV $e^+e^-$ or $\mu^+\mu^-$ colliders are constructed).
For masses of order 1 TeV and an integrated luminosity of 100 $fb^{-1}$, 
we estimate the yield of color triplet vector-like fermion pairs at the LHC 
to be of order $10^4$ events, where they are produced by a combination of 
$gg$ and $q\bar q$ fusion. If the 
$W$ and $Z$ final states produced in the vector-like fermion decays can be 
triggered on with reasonable efficiency this implies that the production of 
such heavy states should be relatively straightforward\cite{tgrold}. The 
production and detection of heavy color-singlet states at a reasonable rate 
seems somewhat more problematic{\cite {dpf}} due to background issues.

\subsection{\it Universality Violations Revisited}

Do the vector-like fermions have visible indirect effects at lower energies? 
We first examine whether the vector-like and SM 
fermion mixing itself induces a sizeable universality violation.  Here, the 
cases where the vector-like fermions are isodoublets or
isosinglets induce quite different effects; recall that in the isodoublet 
vector-like fermion scenario, $\theta_R \sim 0.05$ and the  difference 
$\theta_L^u-\theta_L^d \sim (0.05)^2$, whereas the reverse is true in the 
case of isosinglet fermions.  We thus find the following shifts in the CKM 
element $|V_{ud}|^2$ in each of the above principle models (to leading order 
in the mixing angles)
\begin{eqnarray}
A & : & -(\theta_L^u-\theta_L^d)^2+(\theta_R^u \theta_R^d)^2\,, \nonumber\\ 
B & : & 0\,, \nonumber\\ 
CD & : & 0\,, \\ 
EF& : & -(\theta_L^u)^2-(\theta_L^d)^2\,. \nonumber
\end{eqnarray}
Clearly the effect is very small in the models with isodoublet quarks
(model A), but can be sizable in the isosinglet quark case (model EF). 
In fact, for model EF we see that the effect 
of mixing is to decrease the value of $|V_{ud}|_{eff}^2$ 
relative to the SM expectation by an amount of order $10^{-3}$; this is at 
the level of current sensitivity, as discussed in the previous section, and is 
comparable to the size of the present difference between experiment and the 
expectations of unitarity. 

The other consequence to notice above is 
the null result in the case of models B and CD.  In these scenarios,
the same mixing that affects nuclear decays also appears in the calculation of 
$\mu$ decay and is therefore absorbed into the definition of $G_F$.  However, 
a residual effect from the mixing will remain in the ratio of widths for 
$\pi \to e\nu$ to $\pi \to \mu \nu$.  This results in another shift in these
models, in addition to that from the leptoquark exchange discussed above, from 
the SM expectation for this ratio by an amount
\begin{eqnarray}
B & : & -(\theta_L^\nu-\theta_L^e)^2-(\theta_R^\nu \theta_R^e)^2\,, \nonumber\\ 
CD& : & -(\theta_L^\nu)^2-(\theta_L^e)^2\,, 
\end{eqnarray}
which is negative and can be sizeable in the case of model CD for 
mixing angles of order 0.05. Experimentally{\cite {pdg}}, the value of this
universality testing ratio to its SM expectation is found to be 
$0.9966\pm 0.0030$; we note that the potential deviation from unity is
comparable to the expectation in model CD. 

Lastly, we note that in model B a right-handed charged current is generated 
for the electron, which 
could in principle be observed in $\mu$ decay if $\nu^c$ appeared 
as missing energy or $p_T$. However, the size of the right-handed amplitude 
generated through this mixing is far too small  (by several orders of 
magnitude) to be detected in the Michel spectrum{\cite {pdg}}.  

\subsection{\it $g-2$ of the Electron and Electron Neutrino}

\begin{figure}[htbp]
\centerline{
\psfig{figure=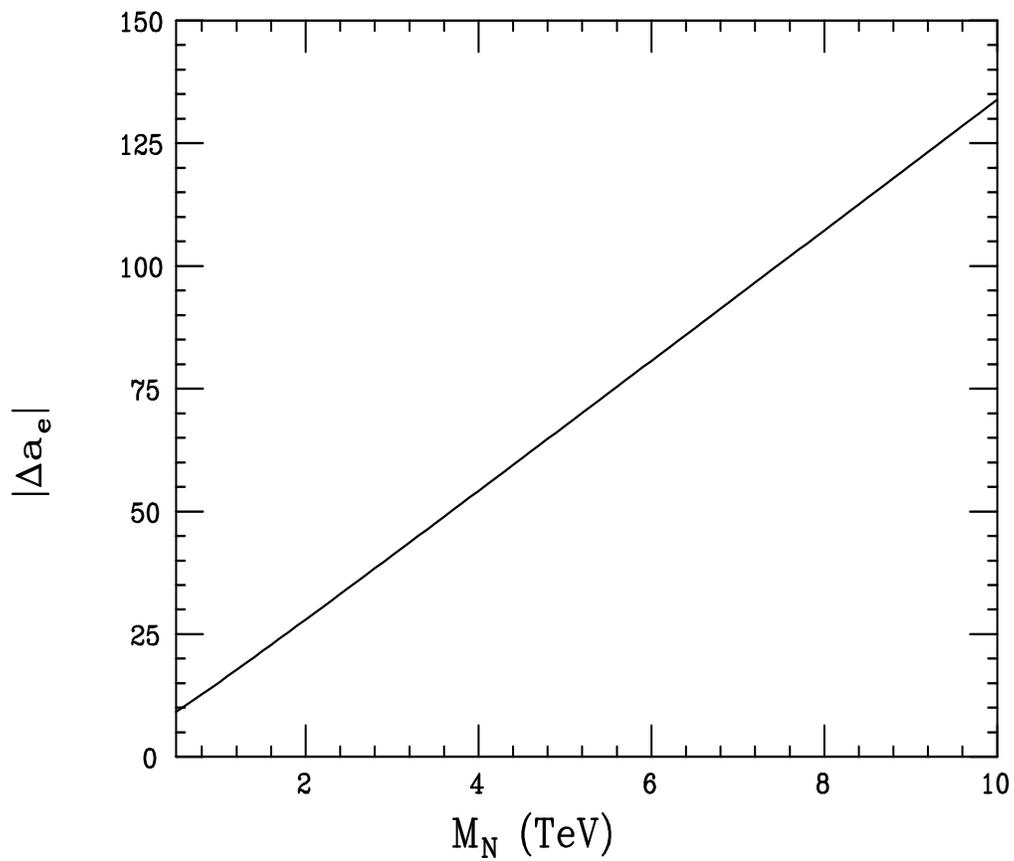,height=14cm,width=16cm,angle=-90}}
\vspace*{-0.9cm}
\caption{ Contribution to the anomalous magnetic moment of the electron in 
model B in units of $10^{-12}$ due to mixing between the SM and the 
vector-like fermions as a function of the $N$ fermion mass.} 
\label{anomc}
\end{figure}

One reason for demanding that the leptoquark 
couplings to fermions be chiral is to 
avoid the enhancement of a number of loop-order processes, \eg, the $g-2$ of 
the electron.  Here we have successfully constructed chirally coupled leptoquark
models and hence their contribution to the electron's $g-2$ is very small, 
however, there remains the possibility that the mixings between the SM 
and vector-like fermions 
may reinstate significant contributions to $a_e$. Model B 
provides an example of this scenario, since in this case, both left- and 
right-handed leptonic couplings to the $W$-boson exist and the heavy $N$
can participate in $a_e$ as an intermediate state.  
The contribution in this case can be immediately obtained 
from Ref.{\cite {lev}} and directly compared with the
prediction of the SM{\cite {dat}} and the experimental value. For the 
difference between the latter two
we find (with the total uncertainty in the difference given in parenthesis)
\begin{equation}
a_e^{exp}-a_e^{SM}=-13.2(27.2)\times 10^{-12}\,,
\end{equation}
while the additional contribution in the case of model B can be written as
\begin{equation}
\Delta a_e=(-34346\times 10^{-12})F(x)\sin(\theta_L^\nu-\theta_L^e)
\sin \theta_R^e\,,
\end{equation}
where $F(x)$ is a kinematical function of the mass ratio $x=M_N^2/M_W^2$. The 
large numerical size of the prefactor gives some warning that the effect 
might be of a reasonable magnitude even though it is highly suppressed by 
several powers of mixing angle factors. 
Taking $\theta_R$ $\sim 0.05$ and the difference $\theta_L^\nu-\theta_L^e
\sim (0.05)^2$ as usual, we obtain the results displayed in 
Fig.\ref{anomc}; note that the absolute value of the shift is presented since 
the signs of the mixing angles are unknown. This analysis demonstrates that 
for typical ranges of the parameters in this model, the size of 
$\Delta a_e$ is comparable to or larger than the present 
uncertainty in the value of $a_e$; hence values of $M_N \gsim 5$ TeV 
are excluded for these suggestive sizes of the mixing angles. 

\vspace*{-0.5cm}
\vspace*{0.4mm}

\begin{figure}[htbp]
\centerline{
\psfig{figure=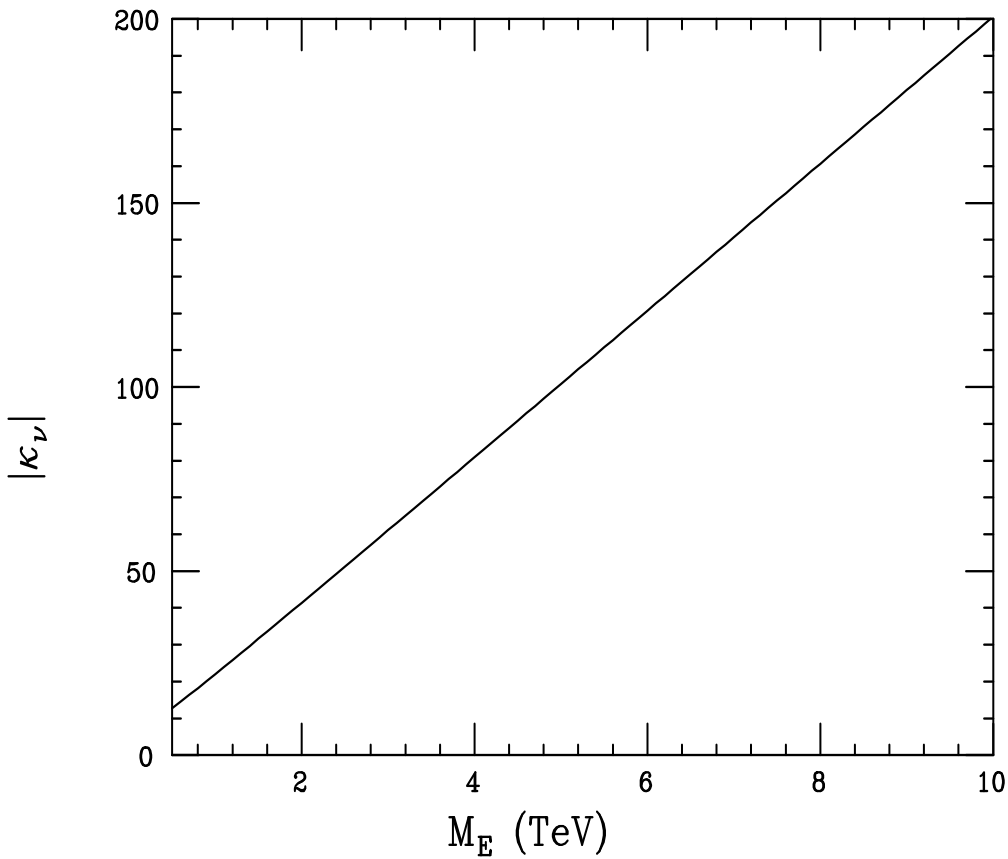,height=14cm,width=16cm,angle=-90}}
\vspace*{-0.9cm}
\caption{Mixing induced contribution to the magnetic moment of the electron 
neutrino in model B in units of $10^{-12}$ Bohr magnetons 
as a function of the $E$ fermion mass.}
\label{anomnu}
\end{figure}

In a similar fashion the corresponding contribution to the magnetic moment of 
the electron neutrino, $\kappa_\nu$, can be obtained,
provided that $\nu_e$ is a Dirac fermion. (We recall that both the electric 
and magnetic dipole moments of a Majorana neutrino vanish identically.) In 
this case the amplitude arises from a penguin diagram with the vector-like 
fermion $E$ in the intermediate state.  The results thus take similar form
to that for $\Delta a_e$ above, except that the 
kinematic function is different and with the replacement 
$\sin\theta_R^e \to \sin\theta_R^\nu$. The result of this calculation is 
presented in Fig.\ref{anomnu} and should be compared to the present experimental
limit{\cite {pdg}} of
$|\kappa_\nu| \leq 180\times 10^{-12} \mu_B$ at $90\%$ CL from elastic 
$\bar \nu_e e$ elastic scattering using reactor neutrinos. Stronger bounds 
(by factors of order 10) based on astrophysical constraints remain somewhat 
controversial{\cite {pdg}}. Note that a similar graph without the attached 
photon is capable of generating a mass for $\nu_e$ in the range
$10^{-3}-10^{-2}$ eV. 

\vspace*{-0.5cm}
\vspace*{0.4mm}

\subsection{\it Oblique Parameters, $Z$ Pole Observables, and APV}

Vector-like fermions are known to have negligible contributions\cite{htt}
to the oblique parameters\cite{obl}.  However, once vector-like fermions 
mix with their SM counterparts it is possible to induce non-zero shifts
in the values of these parameters.  As a numerical example,
we examine the size of these contributions in model A.
In the case of the shift in the $\rho$ parameter, $\Delta\rho
\equiv\alpha T$, there are two sources which contribute here: (i) the 
modification of both the vector-like and SM fermion
couplings to the $W$ and $Z$ due to mixing; (ii)
the $U$ and $D$ masses, originally degenerate, 
are now split by an amount $M_U^2-M_D^2=a_u^2v^2-a_d^2v^{c2}$.  Writing
in this case 
$c_{u,d}=\cos\theta_R^{u,d}$ and $s_{u,d}=\sin\theta_R^{u,d}$, one obtains 
\begin{equation}
\Delta\rho={3G_F\over 8\sqrt 2\pi^2}\left\{ c_u^4M_U^2+c_d^4M_D^2-
2c_u^2c_d^2{M_U^2M_D^2\over M_U^2-M_D^2}\ln {M_U^2\over M_D^2} \right\} \,.
\end{equation}
Note that when $c_u=c_d=1$, $M_U=M_D$ and $\Delta\rho$ vanishes as
expected.  For $M_{U,D}\simeq 5$ TeV and $\theta_R^{u,d}\simeq 0.05$, 
$\Delta\rho$ is found to be $<10^{-4}$, far too small to be observed.  In a
similar vein, the induced value for the parameter $S$ is found to be 
less than $5\times 10^{-4}$ and is hence also vanishingly small. Thus,
although the vector-like fermions 
do not remain purely vector-like after mixing, their
contribution to the oblique parameters remain negligible. 
This same pattern is repeated in the case of the other models with only minor 
differences, \eg, color factors are present in the case of model B and 
the gauge invariant mass terms for the two isosinglet fields in either 
models CD or EF can be different. Numerically, however, similarly small 
results are obtained for the oblique parameters in these remaining models.

Are there observable modifications in the SM fermion couplings to the 
$Z$-boson?  Recall that, for example, the mixing 
of the $u$ and $d$ quarks with vector-like fermions 
which have weak isospin $T_{3u,3d}'$ 
produces a shift in the $u$ and $d$ couplings to the $Z$ of 
$\Delta v_u(a_u)=(T_{3u}'-1/2)(s^u_L)^2\pm T_{3u}'(s^u_R)^2$ 
and $\Delta v_d(a_d)=(T_{3d}'+1/2)(s^d_L)^2\pm T_{3d}'(s^d_R)^2$, respectively, 
using the notation above. (The corresponding shifts in the case of leptonic 
mixing can be obtained from these expressions by trivial notational changes.)
Two places where these coupling shifts may show up most clearly are in the 
partial widths of the $Z$-boson and in APV. In both these observables 
there is an
additional shift in the case of the leptonic couplings due to the overall 
change in the coupling normalization from the redefinition of $G_F$ from 
muon decay, as discussed above.  However, the $Z$ leptonic
asymmetries, which are particularly important observables, are insensitive to 
these overall changes in the coupling normalization. For this case, taking the 
relevant mixing angle to be 0.05 as usual, we find that the $Z$ partial width 
to the $e^+e^-$ final state is decreased(increased) by an amount of order 
$\simeq 0.2$ MeV for the isodoublet(isosinglet) model. 
Correspondingly, the apparent shift in the value of $\sin ^2 \theta^{eff}_w$ 
from the asymmetries increases(decreases) by an amount of order $\simeq 0.0006$ 
for these same two cases. Both of these shifts are essentially at the boundary 
of the current level of sensitivity for LEP/SLD measurements{\cite {moriond}}. 
Similarly there is a corresponding shift in the number of neutrinos extracted 
from the measurement of the $Z$ invisible width by $\simeq 0.005$. 

These shifts in the SM fermion couplings can modify the expectations for APV as 
well since the effective weak charge, $Q_w$, directly probes the two products 
$a_ev_u$ and $a_ev_d$ in addition to the shift in the overall normalization 
that occurs with leptonic mixing. 
In the case where the SM fermions mix with their leptonic vector-like 
counterparts, the shift in $Q_w$ is directly given by 
\begin{equation}
\Delta Q_w/Q_w = \delta \rho-2(T_{3e}'+1/2)(s^e_L)^2+2T_{3e}'(s^e_R)^2\,,
\end{equation}
where $\delta \rho$ represents the change in the overall coupling 
normalization and is given to leading order in the mixing angles by 
\begin{eqnarray}
B & : & (\theta_L^\nu-\theta_L^e)^2-(\theta_R^\nu \theta_R^e)^2\,, \nonumber\\ 
CD& : & (\theta_L^\nu)^2+(\theta_L^e)^2\,.
\end{eqnarray}
We find that this fractional shift in $Q_w$ is at the $10^{-3}$ 
level for either isosinglet or isodoublet leptonic 
vector-like fermions and is hence
clearly too small to be observed. The modification could be potentially larger 
when mixing occurs in the $u$ and $d$ couplings and where there is no overall 
change in the normalization.  However we find that the individual contributions
of the $u$ and $d$ quarks tend to cancel each other instead of adding 
coherently, leaving, again, a relative shift in $Q_w$ at the $10^{-3}$ level. 

\subsection{\it Drell-Yan Production in the $e^\pm \nu_e$ Channel}

What future constraints can be placed on the leptoquark couplings?  We know 
from earlier work{\cite {old}} that the $\lambda_d$ coupling can be probed in 
high precision measurements at LEP II in $e^+e^- \to q \bar q$ and also at
the Tevatron via NC Drell-Yan production.  Can future colliders also probe the 
$\lambda_u$ coupling? One possibility is to examine the corresponding CC 
Drell-Yan process at hadron colliders, $p^(\bar p^)\to e^\pm\nu$.  In addition
to the usual SM $W$-boson exchange, leptoquarks can also contribute to this 
process via t-channel exchange involving both the
$\lambda_d$ and $\lambda_u$ couplings.
The subprocess cross section for this reaction is found to be 
\begin{equation}
{d\hat\sigma(\bar ud\to e^-\bar\nu_e)\over {dz}} =
{G_F^2M_W^4\over {12\pi \hat s}}\left[{\hat u^2\over 
{(\hat s-M_W^2)^2+(\Gamma_W M_W)^2}}+\left({\tilde \lambda_u \tilde \lambda_d
\over {x}}\right)^2{\hat t^2\over {(\hat t-m_{LQ}^2)^2}}\right]\,,
\end{equation}
where $x=G_FM_W^2/2\sqrt 2 \pi \alpha$ and $z=\cos \theta^*$, the parton 
center of mass scattering angle between the incoming quark and the outgoing 
negatively charged electron; as usual $\hat t=-\hat s(1-z)/2$ and 
$\hat u=-\hat s(1+z)/2$. 
Note that there is no interference 
between the $W$-boson and leptoquark exchanges which will make the 
leptoquark contribution somewhat 
more difficult to observe although the two distributions peak in opposite 
angular regions.

\vspace*{-0.5cm}
\nn
\begin{figure}[htbp]
\centerline{
\psfig{figure=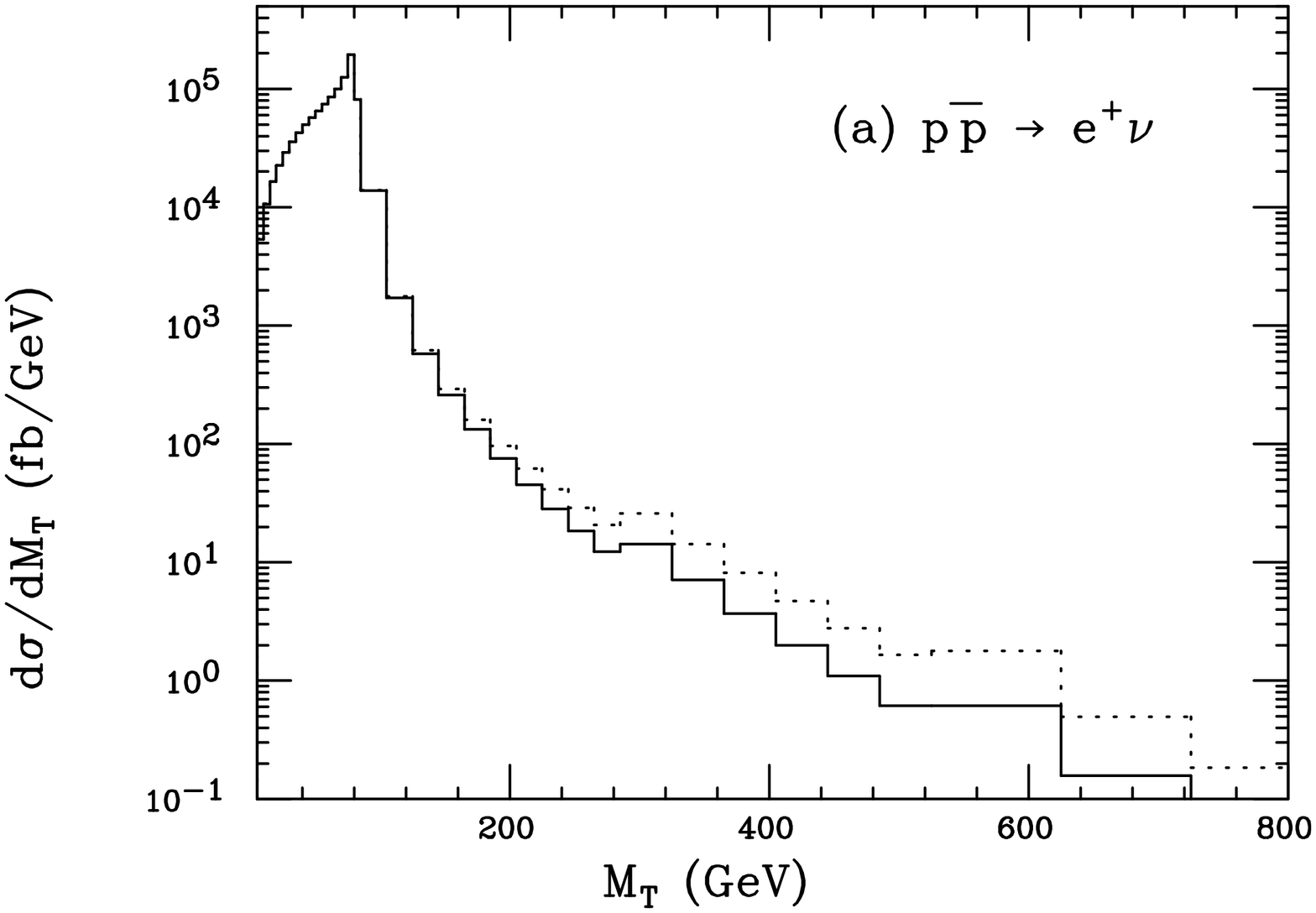,height=10.cm,width=12cm,angle=-90}}
\vspace*{-5mm}
\centerline{
\psfig{figure=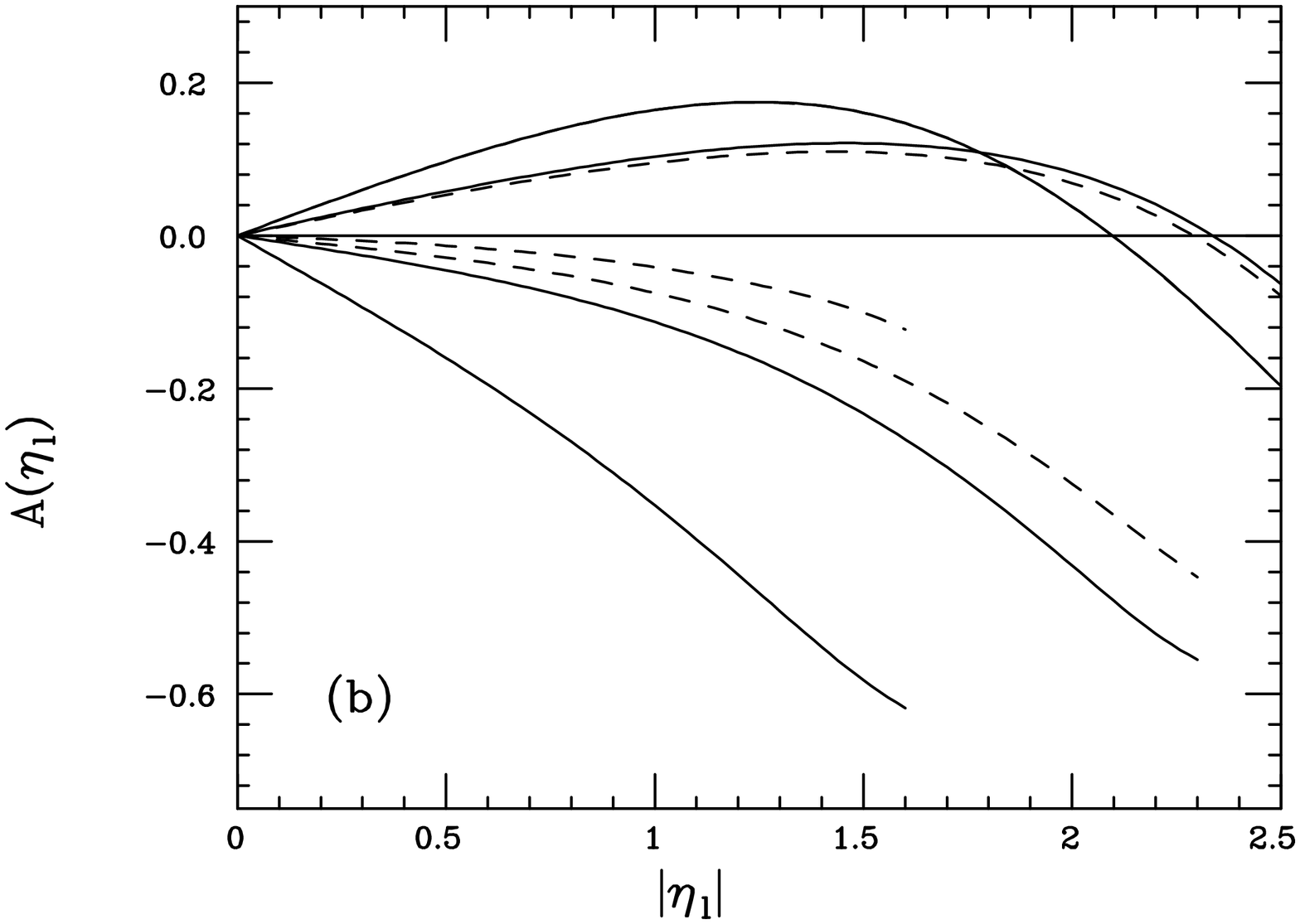,height=10.cm,width=12cm,angle=-90}}
\vspace*{-0.75cm}
\caption{(a) The electron plus neutrino transverse mass distribution 
assuming $|\eta_\ell|\leq 2.5$ and (b) the folded 
lepton charge asymmetry in the charged current Drell-Yan production 
channel at the 2 TeV Tevatron for the SM (solid curves)and with 200 GeV scalar 
leptoquark exchange assuming $\tilde \lambda_u \tilde \lambda_d=1$(dashed 
curves). In (b), from top to bottom in the center of the figure, the SM curves 
correspond to $M_T$ bins of 50-100, 100-200, 200-400 and $>400$ GeV, 
respectively. Note that for $M_T$ in the 50-100 GeV range there is no 
distinction between the SM result and that with a leptoquark.}
\label{joanne1}
\end{figure}

There are two useful observables in this case.  First, one 
can examine the transverse mass ($M_T$) distribution beyond the Jacobian 
peak associated with $W$-boson production.  For large values of $M_T$ one 
would expect an increase in $d\sigma/dM_T$ due to the leptoquark exchange.
A second possibility is to examine the leptonic charge asymmetry, 
$A(\eta_\ell)$, for the case of electrons in the final state as a 
function of their rapidity. Here $A(\eta_\ell)$ is defined as 
\begin{equation}
A(\eta_\ell)={dN_+/d\eta_\ell -dN_-/d\eta_\ell \over {dN_+/d\eta_\ell 
+dN_-/d\eta_\ell}}\,,
\end{equation}
where $N_{\pm}$ are the number of positively/negatively charged electrons of a 
given rapidity.  In the SM, the charge asymmetry is sensitive to the ratio
of u-quark to d-quark parton densities and the $v-a$ structure of the $W$
decay\cite{chargew}.  Since the decay structure of the $W$ has been 
well-measured elsewhere\cite{wdecay}, any observed deviations from SM 
expectations in this asymmetry  have been
attributed to modifications in the parton density functions\cite{mrs}.  The
possibility of new physics contributing to this channel has been overlooked.
In calculating the asymmetry it is essential to split the integration over the
parton densities into 2 regions, corresponding to positive and negative
lepton rapidities in the $W$ center of mass frame, according to the prescription
in Ref. \cite{oldaltar}.
Fig.\ref{joanne1} shows how both the binned transverse mass 
distribution and the lepton charge asymmetry, for four $M_T$ bins corresponding
to $50<M_T<100$ GeV, $100<M_T<200$ GeV, $200<M_T<400$ GeV, and $400<M_T$ GeV,
are modified by 
the presence of a 200 GeV leptoquark with, for purposes of demonstration, 
$\tilde \lambda_u \tilde \lambda_d=1$.  We see that the transverse mass
distribution does rise above the SM expectations for large values of $M_T$ 
as expected, and that the lepton charge asymmetry can also be significantly
modified for larger values of $M_T$.  Note that there is little deviation
in the asymmetry in the transverse mass bin associated with the $W$ peak,
$50<M_T<100$ GeV, so that this $M_T$ region can still be used for determination
of the quark densities.

\begin{figure}[htbp]
\centerline{
\psfig{figure=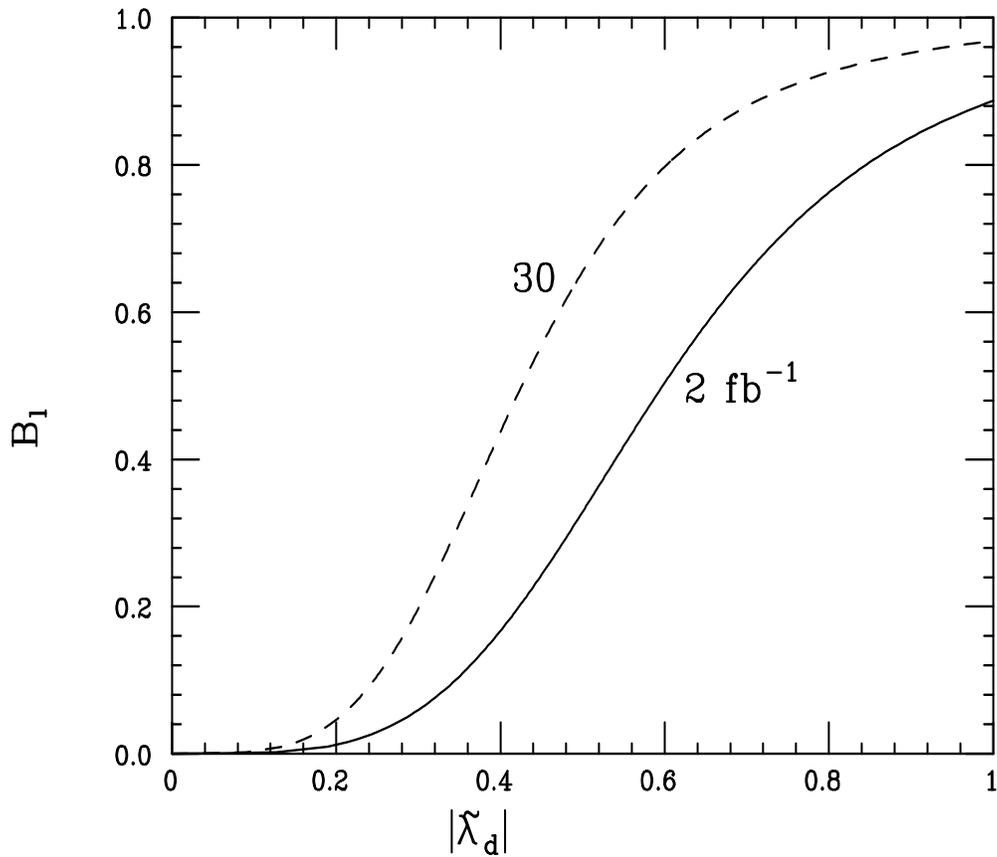,height=14cm,width=16cm,angle=-90}}
\vspace*{-0.9cm}
\caption{$95\%$ CL bound on $B_\ell$ as a function of $\tilde \lambda_d$ 
from a fit to both the $M_T$ distribution and $A(\eta_\ell)$ at the 2 TeV 
Tevatron for two integrated luminosities as indicated. The area below and to 
the right of the curves are excluded.}
\label{joanne2}
\end{figure}

We now perform a $\chi^2$ analysis to determine the potential sensitivity to 
leptoquark exchange at the main injector.  As shown in Fig.\ref{joanne1}(a),
we divide the transverse mass distribution into several bins corresponding to
\begin{eqnarray}
{\mbox{7 bins in steps of 5 GeV in the range}} & 50\leq M_T\leq 85 & 
{\mbox{GeV}}\,, \nonumber\\
{\mbox{10 bins in steps of 20 GeV in the range}} & 85\leq M_T\leq 285 & 
{\mbox{GeV}}\,, \nonumber\\
{\mbox{6 bins in steps of 40 GeV in the range}} & 285\leq M_T\leq 525 & 
{\mbox{GeV}}\,, \\
{\mbox{2 bins in steps of 100 GeV in the range}} & 525\leq M_T\leq 725 & 
{\mbox{GeV}}\,, \nonumber\\
{\mbox{1 bin for the range}} & 725\leq M_T & {\mbox{GeV}}\,. \nonumber
\end{eqnarray}
This ensures that adequate statistics are maintained in each bin.  The
apparent rise in the cross section in Fig. \ref{joanne1}(a) at $M_T=285$
GeV is an artifact of the increased bin width at that point.  For the
lepton charge asymmetry the lepton's rapidity is binned as
\begin{eqnarray}
{\mbox{12 bins in steps of $\Delta\eta_\ell=0.2$ in the range}} & 
50\leq M_T\leq 100 & {\mbox{GeV}}\,, \nonumber\\
{\mbox{12 bins in steps of $\Delta\eta_\ell=0.2$ in the range}} & 
100\leq M_T\leq 200 & {\mbox{GeV}}\,, \\
{\mbox{6 bins in steps of $\Delta\eta_\ell=0.3$ in the range}} & 
200\leq M_T\leq 400 & {\mbox{GeV}}\,, \nonumber\\
{\mbox{2 bins of $|\Delta\eta_\ell|\leq 0.4$ and 
$|\Delta\eta_\ell|\geq 0.4$ in the range}} & 
400\leq M_T & {\mbox{GeV}}\,, \nonumber
\end{eqnarray}
subject, of course, to the constraint $M_T\leq\sqrt se^{-|\eta_\ell|}$.
We note again, that there is no sensitivity to the leptoquark exchange on
the $W$ transverse mass peak ($50<M_T<100$ GeV), and hence this region does
not contribute to the $\chi^2$ distribution.  The bin integrated cross
section and asymmetry are then obtained for the SM and for the case of
200 GeV scalar leptoquark exchange.  We sum over both $e^+\nu_e$ and
$e^-\bar\nu_e$ production for the cross section and employ an electron 
identification efficiency of $0.75$.  The statistical errors are
evaluated as $\delta N=\sqrt N$ and $\delta A =\sqrt{(1-A^2)/N}$, as usual.
The resulting $95\%$ C.L. bound in the $B_\ell - \tilde\lambda_d$ plane
is presented in Fig. \ref{joanne2} for 2 and 30 fb$^{-1}$ of integrated
luminosity with $\sqrt s=2$ TeV.  We see that even for 30 fb$^{-1}$, the
constraints are inferior to those obtained from present data on $\pi$ decay
universality as shown in Fig. 2.

\vspace*{-0.5cm}
\vspace*{0.4mm}

\subsection{\it Like-Sign Leptoquark Production at the Tevatron}

\begin{figure}[htbp]
\centerline{
\psfig{figure=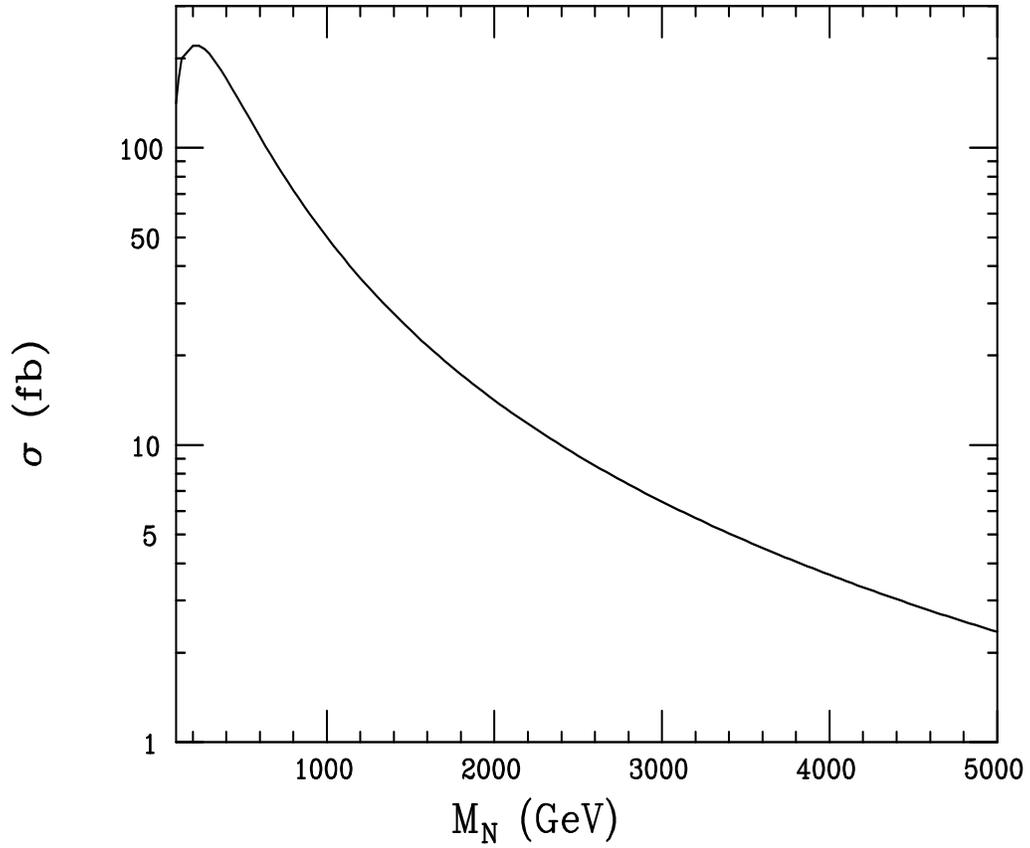,height=14cm,width=16cm,angle=-90}}
\vspace*{-0.9cm}
\caption{ Cross section for like-sign pairs of 200 GeV leptoquarks at the 2 TeV 
Tevatron as a function of the neutral vector-like fermion mass, 
$M_N$. The Yukawa coupling is assumed to be unity.}
\label{crazy}
\end{figure}

In models B and C where the $u$ quark couples to a heavy neutral vector-like 
fermion, $N$, 
new processes may arise if $N$ is a Majorana field. (Note that for simplicity 
we have only considered the Dirac case in the above discussions.) One such 
unusual possibility is the production of 
pairs of {\it identical} leptoquarks in hadronic collisions via 
$u-$ or $t-$channel $N$ exchange which generates the process $uu\to 2LQ$. The 
leptoquarks then decay to like-sign charged 
leptons plus jets, a relatively clean signature at a hadron collider. 
Recall that the relevant Yukawa coupling involved in 
this $\Delta L=2$ reaction is of order unity so that this cross section may be 
significant even though it is a valence times sea-quark density process at the 
Tevatron.  We find the subprocess cross section to be
\begin{equation}
{d\sigma\over {dz}} ={\lambda^4\over {128\pi}}\beta 
\left[{M_N(\hat t+\hat u-2M_N^2)
\over {(\hat t-M_N^2)(\hat u-M_N^2)}}\right]^2\,,
\end{equation}
where $\lambda \sim 1$, $z$ is defined in the previous section, and $M_N$ is 
the mass of the neutral vector-like fermion. Here, $\hat t=-\hat s(1-
\beta z)/2$ and $\hat u=-\hat s(1+\beta z)/2$, where $\beta=(1-4m_{LQ}^2/\hat s
)^{1/2}$. Note that 
as $M_N \to 0$ the cross section vanishes as expected for a Majorana 
fermion induced process. The rate for this reaction at the Tevatron with 
$\sqrt s$=2 TeV is shown in Fig.\ref{crazy} for $\lambda=1$. Here we see that 
the cross section initially rises with increasing $M_N$ but then begins to fall,
scaling like $M_N^{-2}$ as $M_N \to \infty$. For $M_N$=1 TeV, 
this cross section corresponds to $\simeq$ 100 events at the Main Injector 
{\it before} leptoquark branching fractions are taken into account. At the 
$\sqrt s$=1.8 TeV collider the cross section is smaller by a factor of 
$\simeq 0.56$. 

\vspace*{-0.5cm}
\vspace*{0.4mm}

\subsection{\it Speculations on a Realistic Flavor Coupling Structure}

Although one can impose discrete or other symmetries so that leptoquarks only 
couple to a single generation in the weak eigenstate basis it is difficult to 
understand how this might hold in the physical basis. This 
issue is a major stumbling block for the construction of realistic leptoquark 
models 
and is one that we have carefully avoided until now. Of course the detailed 
exploration of possible solutions to this problem lies outside the scope of 
this paper{\cite {tj}}, however, there are directions that do show some 
promise{\cite {leurer}}. 

To be more specific, let us concentrate on models which yield 
the interaction ${\cal L}_{wanted}$ in Eq. (1) where the $\nu^c$ field is 
absent, and also
neglect the possibility of any large leptonic mixing. This implies that all of 
the traditional flavor changing neutral current (FCNC) constraints are only to 
be applied to the quark sector of the model as lepton number is conserved. 
Interestingly, in this case the relevant flavor changing terms are induced by 
the right-handed unitary matrices $U(u,d)_R$ of which there is no information
since they play no role in SM interactions.  For purposes of demonstration
we will this simply assume that element for element they are numerically 
similar to the corresponding CKM matrix elements. Thus flavor mixing now leads 
us to make the substitutions 
$u_R \to \sum_i [U(u)_R]_{1i}(u_R)_i$ and similarly for $d_R$ in 
${\cal L}_{wanted}$. This particular form guarantees that tree level
$s\to d$ and $b\to d,s$ transitions 
will be accompanied by $e^+e^-$, while $c\to u$ processes are accompanied only 
by $\bar \nu_e \nu_e$.  Thus leptoquarks 
will not mediate the potentially dangerous processes 
$K \to \pi \bar \nu \nu$ or $D\to \pi e^+e^-$ at tree level.

Are the tree-level rates induced by these leptoquarks dangerously large? 
Fortunately, the chirality of the leptoquark couplings automatically reduces 
the size of their potential 
contributions to rare processes and the fundamental couplings present in 
the Lagrangian are already quite small. As we saw from our discussion 
of $|V_{ud}|$, for example, the effective Fermi coupling for leptoquark 
exchange was below the level of $10^{-3}G_F$ for typical values of the 
Yukawa couplings. In fact,
using $\tilde \lambda_{u,d} \simeq 0.15$ and $M_{LQ}=200$ GeV it is easy to 
show that this class of models indeed satisfies all of the FCNC experimental 
constraints in Ref. {\cite {leurer}} for values 
of $[U(d)_R]_{sd},[U(u)_R]_{uc}$ of order 0.1-0.2. (This result remains true 
even when these constraints are strengthened by more recent experimental 
results{\cite {pdg}}). This observation leads us to believe that leptoquarks of 
the type under discussion here are not only compatible with present bounds 
from flavor changing data but may lead to new effects in flavor physics that 
are comparable 
in magnitude to SM contributions and can thus be searched for in charm or B 
factories{\cite {tj}}. 

\section{Unification? Never Break the Chain}

\subsection{\it Non-SUSY Case}

If leptoquarks exist and we also believe that there is experimental 
evidence for coupling constant unification then we must begin to examine 
schemes which contain both ingredients as pointed out in our 
earlier work{\cite {old}}. In the scenarios 
at hand, the SM quantum numbers of the leptoquark are fixed but new vector-like 
fermions have now been introduced as well, all of which will alter the usual 
RGE analysis of the running couplings. 

\begin{figure}[htbp]
\centerline{
\psfig{figure=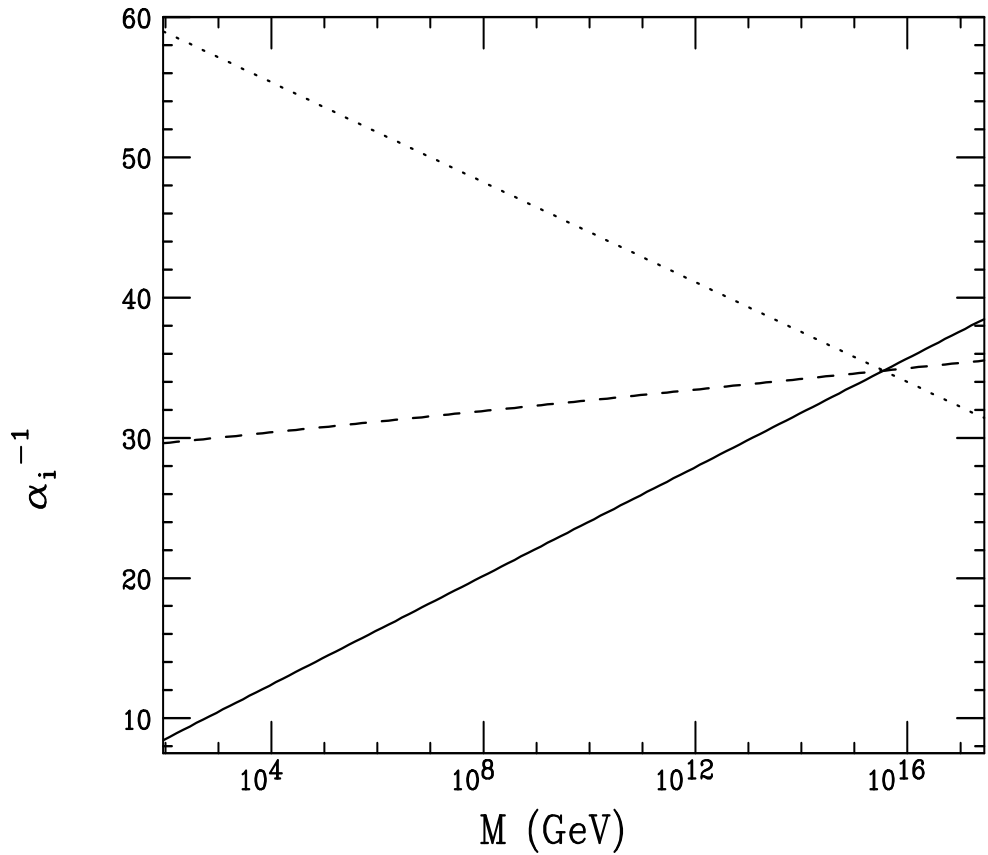,height=14cm,width=16cm,angle=-90}}
\vspace*{-0.9cm}
\caption{Two-loop RGE evolution of the model with the SM particle content 
together with a leptoquark and its conjugate as well as with the vector-like 
fermions and 
Higgs content of model A. The $SU(3)_C(SU(2)_L,~U(1)_Y)$ coupling corresponds 
to the dotted(dashed, solid) curve.}
\label{guts}
\end{figure}

Before discussing supersymmetric models we note that coupling constant
unification {\it can} occur in leptoquark 
models containing exotic fermions even if SUSY is not introduced as was 
shown many years ago in Refs. {\cite {my,oldt}}. Of 
course in the work of Murayama and Yanagida{\cite {my}}, the leptoquark was an 
isodoublet which was one of the BRW models, and is now excluded by 
the combined HERA and Tevatron data. In the scenarios presented here the 
leptoquark is now a $Q=2/3$ isosinglet so 
that the Murayama and Yanagida analysis does 
not apply. Fortunately, we see from the results of Ref. {\cite {oldt}} that a 
second unification possibility does exist for just these types of models: 
in addition to the SM spectrum, one adds a
leptoquark and its conjugate as well as a vector-like pair of 
color-triplet, 
isodoublets together with the field $H^c$. This is just the particle 
content of model A. To verify and update this earlier analysis, we 
assume for simplicity that all the new matter fields 
are introduced at the weak scale and take $\sin ^2 \theta_w=0.2315$ as input 
to a two-loop RGE study. The results are shown in Fig.\ref{guts} 
where we obtain the 
predictions that coupling unification occurs at $3.5 \times 10^{15}$ GeV and 
$\alpha_s(M_Z)$ is predicted to be 0.118. If unification does indeed occur we 
can estimate the proton lifetime{\cite {mrbill}} to be 
$\tau_p=1.6\times 10^{34\pm 1}$ years, safely above current 
constraints{\cite {pdg}}. We find this situation to be intriguing and we 
leave it to the reader to ponder further. 

\vspace*{-0.5cm}
\vspace*{0.4mm}

\subsection{\it SUSY Models}

Of course there are other reasons to introduce SUSY beyond that of 
coupling constant unification, so we now turn to the SUSY versions of 
the above leptoquark models with the assumption that $R$-parity is 
preserved, \ie, the HERA excess is due to a leptoquark and not a 
squark produced through R-parity violating interactions.
This subject was discussed at some length in our earlier 
work{\cite {old}} from a somewhat different viewpoint but from which we are 
reminded of several important points:

($i$) To {\it trivially} preserve the 
successful unification of the SUSY-SM, only complete $SU(5)$ representations 
can be added to the conventional MSSM spectrum. As is well-known, the 
addition of extra matter superfields in complete $SU(5)$ representations 
delays unification 
and brings the GUT scale much closer to the string scale. Of course, 
there still remains the rather unnatural possibility of adding incomplete, but 
`wisely chosen', split representations. This is what happens, of course, in 
the case of the usual Higgs doublets and is the basis for the famous 
doublet-triplet splitting problem. Employing split representations 
certainly allows more 
flexibility at the price of naturalness but still requires one to choose 
sets of $SU(3)_C\times SU(2)_L \times U(1)_Y$ representations 
which will maintain asymptotic freedom and perturbative unification. 
Of course one would still need to eventually explain why these multiplets 
were split. An example of this rather bizarre scenario is the possibility of 
adding a $(2,3)(1/6)$ from a {\bf 15} and a
$(1,1)(1)\oplus (1,\bar 3)(-2/3)$ from a {\bf 10} to the spectrum 
at low energy{\cite {old}}. Here the notation 
refers to the $(SU(3)_C,SU(2)_L)(Y/2)$ quantum numbers of the representation. 
We remind the reader that in this notation the leptoquark itself transforms as 
$(1,3)(2/3)$; the smallest standard $SU(5)$ representation into which the 
${\mbox{LQ}}+{\mbox{LQ}}^c$  
can be embedded is a {\bf 10}$\oplus\overline{\mbox{\bf 10}}$, while in 
flipped-$SU(5)\times U(1)${\cite {flip}}, it can be placed in a 
{\bf 5}$\oplus\overline{\mbox{\bf 5}}$. 

($ii$) Since we only have vector-like fermions in our models, 
it is clear that 
only pairs of representations,  {\bf R}$+\overline{\mbox{\bf R}}$, can be 
added to the MSSM spectrum in order to maintain anomaly cancelation. Of 
course this is also true for the leptoquark superfield in that 
both the LQ and LQ$^c$ fields must now be present as discussed above. 

($iii$) To preserve perturbation 
theory and asymptotic freedom up to the GUT scale when adding complete 
representations, at most one {\bf 10}$+\overline{\mbox{\bf 10}}$ pair or 
three {\bf 5}$+\overline{\mbox{\bf 5}}$ pairs can be appended to the low energy 
spectrum of the MSSM apart from SM singlets. The reason for this is the 
general observation that if one adds more than 
three, vector-like, color triplet superfields to the MSSM particle content then 
the one-loop QCD beta function changes sign. Recall that the leptoquark itself 
already accounts for one of these color triplets. This same 
consideration also excludes the introduction of light exotic fields in 
higher dimensional $SU(3)_C$ representations. Complete $SU(5)$ 
representations larger than {\bf 10}$+\overline{\mbox{\bf 10}}$ are found to 
contribute more than this critical amount to the running of the QCD coupling 
which would then blow up long before the GUT scale is reached. Whether 
unification with strong coupling is possible has been considered 
elsewhere{\cite {ross}}, but we disregard this possibility here.

These are highly restrictive constraints on the 
construction of a successful GUT scenario containing both vector-like fermions 
and leptoquarks and we see that none of the models discussed above can 
immediately satisfy them {\it unless} the leptoquark and vector-like fermion
superfields can be placed into a single $SU(5)$ representation. In the 
standard $SU(5)$ picture, we can then place $(U,D)^T$, an isosinglet $E^c$ and 
$LQ^c$ into a single {\bf 10} with the corresponding conjugate fields in the 
$\overline{\mbox{\bf 10}}$. This would form a hybrid of model A with the 
`skeleton' model D, which we have denoted by AD in Table 1. Of course we pay no 
penalty for also 
including the `skeleton' model C here as well, which then yields model ADC.
Instead, when we consider the flipped-$SU(5)\times U(1)$ case, it would appear 
that we can place 
$(N,E)^T$ and $LQ^c$ into a $\overline{\mbox{\bf 5}}$ with the conjugate 
fields in the {\bf 5}; this is exactly model B. It would also seem that 
no penalty is paid as far as unification is concerned for including the 
`skeleton' model C here as well  {\it except} that this would violate our
assumption about the chirality of 
leptoquark couplings to fermions. However, this model is no 
longer truly unified since the hypercharge generator is not fully contained 
within the $SU(5)$ group itself and lies partly in the additional $U(1)$. 
While the $SU(3)_C$ and $SU(2)_L$ couplings will unify, $U(1)_Y$ will not join 
them even when arbitrary additional vector-like singlet fields are added. 
Thus unification no longer occurs in this scenario so that this possibility 
is now excluded. 

The leptoquark embedding situation becomes more perplexing if the leptoquark 
and vector-like fermions cannot 
occupy the same GUT multiplet. In this case unification and asymptotic 
freedom constraints become particularly tight and we are forced to consider 
the split multiplet approach mentioned above. This means that we add the 
fields $(2,3)(1/6)\oplus (1,1)(1)\oplus (1,\bar 3)(-2/3)$ and 
their conjugates at low energies but constrain them to be from different 
$SU(5)$ representations. In this case the combination 
$(1,3)(2/3)\oplus (1,\bar 3)(-2/3)$ corresponds to the isosinglet leptoquark 
and its conjugate so what remains can only be the 
vector-like fermion fields. Note that we have again 
arrived back at models AD and ADC. Are these the only solutions? We have 
performed a systematic scan over a very large set of 
vector-like fermions with various 
electroweak quantum numbers under the assumption that they are either color 
singlets or triplets, demanding only that ($i$) QCD remains asymptotically 
free and ($ii$) the model passes the so-called ``B-test''{\cite {mpeskin}} 
which is highly non-trivial to arrange. Essentially the B-test takes advantage 
of the observation that if we know the couplings at the weak scale and we 
demand that unification takes place {\it somewhere} then the values of the 
one-loop beta functions must be related. Note that it is a necessary but not 
sufficient test on our choice of models but is very useful at chopping away 
a large region of parameter space. Using the latest experimental 
data{\cite {moriond}}, we find that
\begin{equation}
B = {b_3-b_2\over {b_2-b_1}} = 0.720\pm 0.030   \,, 
\end{equation}
where the $\pm 0.030$ is an estimate of the corrections due to higher order as 
well as threshold effects and the $b_i$ are the one-loop beta functions of the 
three SM gauge groups. Note that $B_{MSSM}=5/7 \simeq 0.714$ clearly satisfies 
the test. If we require that ($i$) and ($ii$) be satisfied and also require 
that the unification scale not be too low then only the solutions described 
above survive after examining $>7\times 10^{7}$ combinations of matter 
representations. While not completely exhaustive, this search indicates the 
solutions above are fairly unique. It is interesting to observe that models 
constructed around model A produce successful grand unification both with and 
without SUSY.

Finally we need to briefly comment on the possible relationship between the 
LQ and LQ$^c$ masses and their SUSY partners. 
In these SUSY models one might imagine that the fermionic partner of 
the leptoquark, the leptoquarkino, may have a mass comparable to the 
vector-like fermions, \ie, of 
order 1-5 TeV or so. Why then is the leptoquark itself so light? One possible 
mechanism, discussed in another context by Deshpande and Dutta{\cite {big}}, 
is to envision a large mixing between the leptoquark 
and its conjugate that produces a
see-saw effect analogous to what happens in light stop quark 
scenarios{\cite {stop}}. This possibility will not be pursued further here. 

\section{Summary and Conclusions}

In this paper, 
we have obtained a general framework for the construction of new $F=0$ scalar 
leptoquark models which go beyond the original classification by Buchm\"uller, 
R\"uckl and Wyler. This approach is based on the observation that in any 
realistic extension of the SM containing leptoquarks it is expected 
that the leptoquarks themselves will 
not be the only new ingredient. This construction technique is, of course, 
far more general than that required to address the specific issue of the HERA 
excess and, as outlined, can also be used to obtain a new class of 
$F=2$ scalar leptoquark models if so desired. 

To extend leptoquark models into new territories it is 
necessary to re-examine the assumptions that have gone into the classic BRW 
framework. While the assumptions of gauge invariance 
and renormalizability are unquestionable requirements of model 
building, it is possible that the other conditions one usually 
imposes are much too strong--unless they are clearly demanded by data. 
This observation implies that for leptoquarks to be experimentally accessible 
now, or anytime soon, their couplings to SM fermions must be essentially 
chiral and separately conserve both Baryon and Lepton numbers. The assumption 
that leptoquarks couple to only a single 
SM generation is surely convenient by way of avoiding numerous low energy 
flavor changing neutral current constraints but is far from natural in the mass 
eigenstate basis. Our analysis indicates that the natural imposition of this 
condition in the original weak basis, and then allowing for CKM-like 
intergenerational mixing does not obviously cause any difficulties 
with experimental constraints, 
especially if lepton generation number is at least approximately conserved. 
What is required to obtain a new class of leptoquark models is that 
the leptoquarks themselves must be 
free to couple to more than just the SM fermions and gauge fields. 

Given the fixed gauge structure of the SM the most likely new interactions 
that leptoquarks may possess are with the Higgs field(s) 
responsible for spontaneous 
symmetry breaking and with new vector-like fermions that are a common 
feature in many extensions of the SM. Such particles have the advantages 
that are automatically anomaly free and give essentially no 
significant contributions to the oblique parameters. 
In the analysis presented above we have 
shown how two particular new forms of the effective interactions of 
leptoquarks with 
the SM fermions, consistent with Tevatron searches, the HERA excess in both 
the NC and CC channels and low-energy data, can arise through the action of 
vector-like fermions and ordinary symmetry breaking.  The typical 
vector-like fermion mass was found 
to lie in the low TeV region and they could thus be directly produced at 
future colliders with known rates. With our set of assumptions, we obtained 
ten new models which fell into two broad classes according to the chirality of 
the resulting leptoquark couplings to the SM fermions. The vector-like 
fermions themselves were shown 
to lead to a number of model-dependent effects which are close to the boundary 
of present experimental sensitivity including ($i$) violations of quark-lepton 
universality (for which, as discussed, there is some evidence at the $2\sigma$ 
level arising from the CKM matrix), 
($ii$) possible small changes in the $Z$-pole observables for 
electrons, ($iii$) a small contribution to the shift in the value of the 
weak charge measured by atomic parity violation experiments over and above 
that induced by the leptoquark itself, ($iv$) a new contribution to 
the anomalous 
magnetic moments of the electron and electron neutrino, and ($v$) the possible 
production of like-sign leptoquarks with a reasonable cross section at the Main 
Injector. We also showed that, as in the case of Drell-Yan in the 
$e^+e^-$ channel at the Tevatron discussed in our earlier work, 
there is some potential sensitivity to $t-$channel leptoquark exchange in the 
corresponding $e^\pm \nu$ channel through the transverse mass distribution 
and the charged lepton asymmetry. 

Leptoquarks within the framework of models containing vector-like fermions 
were shown to be consistent 
with Grand Unification in both a supersymmetric {\it and}~non-supersymmetric 
context. The common feature of both schemes is the structure associated with 
model A, \ie, the vector-like fermions are color triplet, weak isodoublets 
in a $(2,3)(1/6)$ 
representation and both $H$ and $H^c$ Higgs fields are required to be 
present as is $LQ^c$ field. In both scenarios the GUT scale is raised 
appreciably from the corresponding model wherein leptoquarks and vector-like 
fermions are absent. In the SUSY case a $(1,1)(1)$ field is also 
required with the optional addition of a SM singlet, corresponding to models 
AD and ACD. In some sense, ACD is the ``anti-$E_6$'' model in that the 
color triplet vector-like fermions are in isodoublets while the color singlet 
fields are all 
isosinglets. Interestingly, in this scenario there is a vector-like 
fermion corresponding to every type of SM fermion. 

Realistic leptoquark models provide a rich source of new physics beyond the
Standard Model.

\noindent{\Large\bf Acknowledgements}

We appreciate comments, discussions and input from 
S. Eno(D0), G. Landsberg(D0), J. Conway(CDF) and H. Frisch(CDF) regarding the 
current Tevatron constraints on the mass of the first generation scalar 
leptoquark. We thank Y. Sirois(H1) and D. Krakauer(ZEUS) for discussions of 
the HERA data.  We also thank H. Dreiner, M. Kr\"amer, R. R\" uckl and 
A. Kagan for informative theoretical discussions as well as N. Glover, T. Han 
and L. Nodulman for informative communications on the lepton rapidity 
asymmetry. 

\newpage

%
%%%%%%%%%%%%%%%%%%--- References
%%%%%%%%%%%%%%%%%%%%%%%%%%%%%%%%%%%%%%%%%%%%%%%%%%%%%%%
\def\MPL #1 #2 #3 {Mod. Phys. Lett. {\bf#1},\ #2 (#3)}
\def\NPB #1 #2 #3 {Nucl. Phys. {\bf#1},\ #2 (#3)}
\def\PLB #1 #2 #3 {Phys. Lett. {\bf#1},\ #2 (#3)}
\def\PR #1 #2 #3 {Phys. Rep. {\bf#1},\ #2 (#3)}
\def\PRD #1 #2 #3 {Phys. Rev. {\bf#1},\ #2 (#3)}
\def\PRL #1 #2 #3 {Phys. Rev. Lett. {\bf#1},\ #2 (#3)}
\def\RMP #1 #2 #3 {Rev. Mod. Phys. {\bf#1},\ #2 (#3)}
\def\ZPC #1 #2 #3 {Z. Phys. {\bf#1},\ #2 (#3)}
\def\IJMP #1 #2 #3 {Int. J. Mod. Phys. {\bf#1},\ #2 (#3)}


\begin{thebibliography}{99}

\bibitem{h1}
C. Adloff \etal, H1 Collaboration, \ZPC C74 191 1997 .
%
\bibitem{zeus}
J. Breitweg \etal, ZEUS Collaboration, \ZPC C74 207 1997 .
%
\bibitem{krak}
D. Krakauer, ZEUS Collaboration, talk given at the {\it $12^{th}$ Workshop on 
Hadron Collider Physics}, Stony Brook, NY June 5-11, 1997. 
%
\bibitem{vb}
For a comprehensive analysis of this possibility and original references, see 
V. Barger \etal, hep-ph/9707412. The analysis by these authors apparently 
excludes this possibility as the source of the HERA excess in the NC channel.
%
\bibitem{parton}
S. Kuhlman, H.L. Lai and W.K. Tung, hep-ph/9704338;
K.S. Babu, C. Kolda and J. March-Russell, hep-ph/9705399;
J.F. Gunion and R. Vogt, hep-ph/9706252;
S. Rock and P. Bosted, hep-ph/9706436;
W. Melnitchouk and A.W. Thomas, hep-ph/9707387. 
%
\bibitem{web}
Whether the excess can be due to a single resonance remains uncertain. It 
appears that the H1 excess is centered at $201\pm 5$ GeV while that for 
ZEUS is at $219\pm 9$ GeV. For overviews, see G. Altarelli, hep-ph/9708437
and hep-ph/9710434; J. Ellis, talk given at the {\it International 
Europhysics Conference on High Energy Physics}, Jerusalem, August 19-26, 1997;
R. R\" uckl and H. Spiesberger, hep-ph/9710327.
%%
\bibitem{big}
D. Choudhury and S. Raychaudhuri, hep-ph/9702392;
G. Altarelli \etal, hep-ph/9703276; 
H. Dreiner and P. Morawitz, hep-ph/9703279;
J. Bl\"umlein, hep-ph/9703287; 
J. Kalinowski \etal, hep-ph/9703288;
K.S. Babu \etal, hep-ph/9703299;
M. Suzuki, hep-ph/9703316;
G.K. Leontaris and J.D. Vergados, hep-ph/9703338;
T. Kon and T. Kobayashi, hep-ph/9704221;
I. Montvay, hep-ph/9704280;
S.F. King and G.K. Leontaris, hep-ph/9704336;
J. Elwood and A. Faraggi, hep-ph/9704363;
B. Dutta, R.N. Mohapatra and S. Nandi, hep-ph/9704428;
M. Heyssler and W.J. Stirling, hep-ph/9705229;
J. Ellis, S. Lola and K. Sridhar, hep-ph/9705416;
S. Jadach, B.F.L. Ward and Z. Was, hep-ph/9705429;
J.E. Kim and P. Ko, hep-ph/9706387;
A. Blumhofer and B. Lampe, hep-ph/9706454;
E. Keith and E. Ma, hep-ph/9707214;
N.G. Deshpande and B. Dutta, hep-ph/9707274
T. Kon \etal, hep-ph/9707355
M. Carena \etal, hep-ph/9707458.
%
\bibitem{old}
J.L Hewett and T.G. Rizzo, \PRD D56 5709 1997 .
%
\bibitem{tev}
J. Hobbs, (D0  Collaboration), presented at the {\it $32^{nd}$ 
Recontres de Moriond: Electroweak Interactions and Unified Theories}, Le Arces, 
France, 15-22 March, 1997;
K. Maeshima (CDF Collaboration), in {\it 28th International Conference on High 
Energy Physics}, Warsaw, Poland, July 1996, FNAL-CONF-96/413-E; F. Abe \etal,
(CDF Collaboration), Fermilab Report FNAL-PUB-96-450-E.
%
\bibitem{brw}
W. Buchm\" uller, R. R\" uckl, and D. Wyler, \PLB B191 442 1987 .
%
\bibitem{rev}
S. Davidson, D. Bailey, and B.A. Campbell, \ZPC C61 613 1994 ;
M. Leurer, \PRD D50 536 1994 , and {\bf D49}, 333 (1994). 
%
\bibitem{cdf}
F. Abe, CDF Collaboration, hep-ex/9708017, submitted to Phys. Rev. Lett. 
See also, H.S Kambara, CDF Collaboration, talk given at the {\it $12^{th}$ 
Workshop on Hadron Collider Physics}, Stony Brook, NY June 5-11, 1997. 
%
\bibitem{d0}
B. Abbott \etal, D0 Collaboration, hep-ex/9707033 and hep-ex/9710032, 
submitted to Phys. Rev. Lett. See 
also, D. Norman, D0 Collaboration, talk given at the {\it $12^{th}$ Workshop on 
Hadron Collider Physics}, Stony Brook, NY June 5-11, 1997. For the latest 
results, see B. Klima, talk given at the {\it International Europhysics 
Conference on High Energy Physics}, Jerusalem, August 19-26, 1997,
hep-ex/9710019.
%
\bibitem{kramer}
M. Kr\"amer, T. Plehn, M. Spira and P. Zerwas, hep-ph/9704322. 
%
\bibitem{babu}
K.S. Babu, C. Kolda and J. March-Russell, hep-ph/9705414. 
%
\bibitem{altar}
G. Altarelli, G.F. Guidice and M.L. Mangano, hep-ph/9705287.
%
\bibitem{htt}
J. Hewett, T. Takeuchi, and S. Thomas, hep-ph/9603391, in {\it Electroweak
Symmetry Breaking and New Physics at the TeV Scale}, ed. T. Barklow \etal,
(World Scientific, Singapore, 1997).
%
\bibitem{obl}
M. Peskin and T. Takeuchi, \PRL 65 964 1990 , and \PRD D46 381 1992 ;
W. Marciano and J. Rosner, \PRL 65 2963 1990 ;
G. Altarelli and R. Barbieri, \PLB B253 161 1990 ;
D. Kennedy and P. Langacker, \PRL 65 2967 1990 , and \PRD D44 1591 1991 ;
I. Maksymyk, C.P. Burgess, and D. London, \PRD D50 529 1994 ;
C.P. Burgess \etal, \PLB B326 276 1994 .
%
\bibitem{wood}
C.S. Wood \etal, Science {\bf 275} 1759 (1997). 
%
\bibitem{ros}
For a recent review and updated estimate, see J.L. Rosner, hep-ph/9704331.
%
\bibitem{pdg}
R.M. Barnett, \etal, (Particle Data Group), \PRD D54 1 1996 .
%
\bibitem{ks}
Z. Kunszt and W.J. Stirling, hep-ph/9703427; 
T. Plehn, H. Spiesberger, M. Spira and P.M. Zerwas, hep-ph/9703433. 
For an updated discussion on the effects of QCD and initial and final state 
radiation, see C. Freiberg, E. Norrbin and T. Sj\"ostrand, hep-ph/9704214. 
%
\bibitem{straub}
B. Straub, ZEUS Collaboration, talk given at the {\it XVIII International 
Symposium on Lepton and Photon Interactions}, July 28-August 1, 1997, Hamburg, 
Germany. For the latest results, see E. Elsen, H1 Collaboration, talk 
given at the {\it International Europhysics 
Conference on High Energy Physics}, Jerusalem, August 19-26, 1997.
%
\bibitem{phyrep}
J.L. Hewett and T.G. Rizzo, \PR 183 193 1989 .
%
\bibitem{ajb}
A. Buras, talk given at the {\it Seventh International Symposium on Heavy 
Flavor Physics}, University of California, Santa Barbara, July 7-11, 1997,
hep-ph/9711217.
This analysis relies heavily on the work presented in E. Hagberg \etal, 
nucl-ex/9609002. 
%
\bibitem{dpf}
See, A. Djouadi, J. Ng, and T.G. Rizzo, in {\it Electroweak Symmetry Breaking
and Beyond the Standard Model}, ed. T. Barklow \etal, (World Scientific,
Singapore), hep-ph/9504210, and references therein; V. Barger, M.S. Berger, and
R.J.N. Phillips, \PRD D52 1663 1995 .
%
\bibitem{tgrold}
T.G. Rizzo, \PLB B181 385 1986 .
%
\bibitem{lev}
J.P. Leveille, \NPB B137 63 1978 .
%
\bibitem{dat}
T. Kinoshita, \PRL 75 4728 1995; J. Ellis \etal, hep-ph/9409376.
%
\bibitem{moriond}
A. Bohm, L3 Collaboration, talk presented at the {\it $32^{nd}$ 
Recontres de Moriond: Electroweak Interactions and Unified Theories}, Le Arces, 
France, 15-22 March, 1997.
%
\bibitem{chargew}
F. Abe \etal, CDF Collaboration, \PRL 74 850 1995 ; 
M.L. Kelley \etal, D0 Collaboration, Fermilab-CONF-96/236-E;
E.L. Berger \etal, \PRD D40 83 1989 ; 
A.D. Martin, R.G. Roberts, and W.J. Stirling, \MPL A4 1135 1989 .
%
\bibitem{wdecay}
G. Arnison \etal, UA1 Collaboration, \PLB 166B 484 1986 ;
R. Ansari \etal, UA2 Collaboration, \PLB 186B 440 1987 ;
B. Balke \etal, TRIUMF Collaboration, \PRD D37 587 1988 .
%
\bibitem{mrs}
H.L. Lai \etal, CTEQ Collaboration, \PRD D51 4763 1995 ; A.D. Martin,
R. G. Roberts, and W.J. Stirling, \PRD D50 6734 1994 .
%
\bibitem{oldaltar}
G. Altarelli \etal, \NPB B246 12 1984 .
%
\bibitem{tj}
J.L. Hewett and T.G. Rizzo, in preparation. 
%
\bibitem{leurer}
For a first discussion of this topic, see M. Leurer in {\cite {rev}}. 
%
\bibitem{my}
H. Murayama and T. Yanagida, \MPL A7 147 1992 .
%
\bibitem{oldt}
T.G. Rizzo, \PRD D45 3903 1992 .
%
\bibitem{mrbill}
W.J. Marciano, talk given at the {\it 1983 International Symposium on
Lepton and Photon Interactions at High Energies}, Ithaca, NY, August 1983.
%
\bibitem{flip}
I. Antoniadis, J. Ellis, J.S. Hagelin and D.V. Nanopolous, 
\PLB B194 321 1987 .
%
\bibitem{ross}
For a recent discussion, see, D. Ghilencea, M. Lanzagorta and G.G. Ross, 
hep-ph/9707462 and references therein.
%
\bibitem{mpeskin}
M. Peskin, SLAC-PUB-7479, 1997.
%
\bibitem{stop}
J.F. Gunion and H.E. Haber, \NPB B272 1 1986 ~and \NPB B402 567E 1993 .
%
\end{thebibliography}
\end{document}